\documentclass[12pt,preprint]{aastex}
\usepackage{natbib}

\newcommand{\caii}{\ion{Ca}{2}}
\newcommand{\etal}{et al.}
\newcommand{\per}{\ensuremath{^{-1}}}

\newcommand{\hal}{H\ensuremath{\alpha}}
\newcommand{\hbe}{H\ensuremath{\beta}}
\newcommand{\hst}{\emph{HST}}
\newcommand{\mage}{MagE}
\newcommand{\esi}{ESI}
\newcommand{\msun}{\ensuremath{M_{\odot}}}
\newcommand{\kms}{km~s\ensuremath{^{-1}}}

\newcommand{\mbh}{\ensuremath{M_\mathrm{BH}}}
\newcommand{\mgal}{\ensuremath{M_\mathrm{gal}}}
\newcommand{\hn}{\hal+[\ion{N}{2}]}
\newcommand{\chisq}{\ensuremath{\chi^2}}
\newcommand{\sigmastar}{\ensuremath{\sigma_{\ast}}}
\newcommand{\mbul}{\ensuremath{\mbh-L_\mathrm{bulge}}}
\newcommand{\msigma}{\ensuremath{\mbh-\sigmastar}}
\newcommand{\rblr}{\ensuremath{R_{\mathrm{BLR}}}}
\newcommand{\fwhm}{\ensuremath{\mathrm{FWHM}}}
\newcommand{\mgb}{\ion{Mg}{1}$b$}
\newcommand{\lbol}{\ensuremath{L_{\mathrm{bol}}}}
\newcommand{\ledd}{\ensuremath{L_{\mathrm{Edd}}}}
\newcommand{\lledd}{\ensuremath{L_{\mathrm{bol}}/L{\mathrm{_{Edd}}}}}
\newcommand{\lhal}{\ensuremath{L_{\mathrm{H{\alpha}}}}}
\newcommand{\lctm}{\ensuremath{L_{5100}}}
\newcommand{\ergs}{ergs~s\ensuremath{^{-1}}}
\newcommand{\at}{\emph{Astronomical Telescope}}
\newcommand{\rl}{\ensuremath{R_{\mathrm{BLR}}-L}}
\newcommand{\magerdc}{\texttt{MAGE\_REDUCE}}

\slugcomment{June 30th 2011}
\shorttitle{LOW-MASS END OF THE \msigma\ RELATION}
\shortauthors{T. Xiao \etal}

\begin{document}
\title{Exploring the Low-Mass End of The \msigma\ Relation with Active Galaxies}
\author {Ting Xiao\altaffilmark{1}, Aaron J. Barth\altaffilmark{2},
Jenny E. Greene\altaffilmark{3}, Luis C. Ho\altaffilmark{4}, Misty
C. Bentz\altaffilmark{5}, Randi R. Ludwig\altaffilmark{3} and Yanfei
Jiang\altaffilmark{6}}

\altaffiltext{1}{Key Laboratory for Research in Galaxies and
Cosmology, Department of Astronomy, University of Science \&
Technology of China, Chinese Academy of Sciences, Hefei, Anhui
230026, China; xiaoting@mail.ustc.edu.cn.}

\altaffiltext{2}{Department of Physics \& Astronomy, University of
California at Irvine, 4129 Frederick Reines Hall, Irvine, CA
92697-4575; barth@uci.edu.}



\altaffiltext{3}{University of Texas at Austin, Department of
Astronomy, 1 University Station, C1400, Austin, Texas 78712-0259,
jgreene@astro.as.utexas.edu}

\altaffiltext{4}{The Observatories of the Carnegie Institution for
Science, 813 Santa Barbara Street, Pasadena, CA 91101;
lho@obs.carnegiescience.edu.}

\altaffiltext{5}{Department of Physics and Astronomy, Georgia State
University, Astronomy Offices, One Park Place South SE, Suite 700,
Atlanta, GA 30303, USA; bentz@chara.gsu.edu}

\altaffiltext{6}{Department of Astrophysical Sciences, Princeton
University, Princeton, NJ 08544, USA; yanfei@astro.princeton.edu}

\begin{abstract}
We present new measurements of stellar velocity dispersions, using
spectra obtained with the Keck Echellette Spectrograph and Imager
(\esi) and the Magellan Echellette (\mage), for 76 Seyfert 1
galaxies from the recent catalogue of Greene \& Ho.  These objects
were selected from the Sloan Digital Sky Survey (SDSS) to have
estimated black hole (BH) masses below $2 \times 10^6$ \msun.
Combining our results with previous ESI observations of similar
objects, we obtain an expanded sample of 93 galaxies and examine the
relation between BH mass and velocity dispersion (the \msigma\
relation) for active galaxies with low BH masses.  The low-mass
active galaxies tend to follow the extrapolation of the \msigma\
relation of inactive galaxies.  Including results for active
galaxies of higher BH mass from the literature, we find a zero point
$\alpha = 7.68 \pm 0.08$ and slope of $\beta = 3.32 \pm 0.22$ for
the \msigma\ relation [in the form $\log \mbh = \alpha + \beta \log
(\sigmastar/200~\mathrm{\kms})]$, with intrinsic scatter of $0.46
\pm 0.03$ dex.  This result is consistent, within the uncertainties,
with the slope of the \msigma\ relation for reverberation-mapped
active galaxies with BH masses from $10^6$ to $10^9$ \msun.  For the
subset of our sample having morphological information from
\emph{Hubble Space Telescope} images, we examine the slope of the
\msigma\ relation separately for subsamples of barred and unbarred
host galaxies, and find no significant evidence for a difference in
slope.  We do find a mild offset between low-inclination and high-inclination
disk galaxies, such that more highly inclined galaxies tend to have
larger \sigmastar\ at a given value of BH mass, presumably due to the
contribution of disk rotation within the spectroscopic aperture.
We also find that the velocity dispersion of the ionized gas, measured
from narrow emission lines including [\ion{N}{2}] $\lambda6583$ \AA, [\ion{S}{2}]
$\lambda\lambda6716,6731$ \AA\AA, and the core of [\ion{O}{3}]
$\lambda5007$ \AA\ (with the blue-shifted wing removed), trace the
stellar velocity dispersion well for this large sample of low-mass
Seyfert 1 galaxies.

\end{abstract}


\section{Introduction}
Dynamical studies of local galaxies over the past decade have firmly
established that supermassive black holes (BHs) are present in most
(possibly all) galaxies with massive bulges, and that the black hole
mass tightly correlates with bulge mass and stellar velocity
dispersion \sigmastar\ \citep{fm00, gebhardt00, mf01, tre02, mh03,
hr04, gult09}.  These correlations suggest coeval growth of the
galaxy bulge and the central BH.  The \msigma\ relation is important
both as a fundamental benchmark against which galaxy evolution
models are tested \citep{hk00}, and as a key input to calculations
of the density of black holes in the Universe \citep{marconi04,
vol08}.

The low-mass end of the mass function locally is where models of
primordial BH seed formation show pronounced differences, thus
improving the observational constraints on the low-mass end of the
\msigma\ relation can provide important constraints on the models.
Theoretical work has proposed two types of seed models: ``light
seeds'' \citep{mr01, vhm03} formed as remnants of Population
\textrm{III} stars, and ``heavy seeds'' \citep{koush04, ln06} from
direct collapse of massive gas clouds in primordial halos.  These
different classes of models should result in different demographics
for BHs in low-mass galaxies at the present epoch.  Low-mass seeds
lead to a wide scatter in BH masses in low-dispersion galaxies (with
some galaxies hosting very low-mass BHs) but a relatively high
occupation fraction of BHs in low-mass galaxies, while in a heavy
seed scenario, the BH occupation fraction is low but the minimum BH
mass is larger \citep{vol08, vol09}.  Unfortunately, observations
have not yet been sufficiently sensitive to distinguish between
these scenarios, and additional observations to constrain the
\msigma\ relation at low masses are needed.  Empirically, the slope
and scatter of the \msigma\ relation are still subject to debate,
particularly at the low and high mass ends \citep{wyithe06}.  There
is some recent work that suggests that the relation may be different
for different types of host galaxies, i.e., for barred vs.\ unbarred
galaxies, and for classical bulges vs.\ pseudo-bulges \citep{hu08,
grali09, gult09, greene10}. These differences may be more pronounced
at lower masses and cause scatter in the overall \msigma\ relation.
However, due to the difficulty of obtaining direct stellar-dynamical
measurements of BH masses in low-mass galaxies, much of the
information on BH demographics at low mass comes from active
galactic nuclei (AGNs).  Recent \emph{Hubble Space Telescope} (\hst)
imaging of low-mass Seyfert 1 galaxies selected from the SDSS by
\citet{GH04} also shows some evidence for a change in the \mbul\
relation at low mass, probably indicative of a different mode of BH
growth in these objects compared with higher-mass galaxies having
classical bulges \citep{ghb08}.  In this paper, our goal is to
expand the sample of low-mass AGNs having both black hole mass
estimates from single-epoch spectroscopy and direct measurements of
stellar velocity dispersion, in order to investigate the low-mass
end of the \msigma\ relation in more detail than was possible
previously.

\citet[][hereafter the GH07 sample]{GH04, GH07c} searched the SDSS
database and presented a large sample of AGNs with low-mass BHs
\citep[$\mbh < 2\times 10^6 \msun$; see also][for related
work]{dong07}. The \sigmastar\ for the GH07 sample could not be
determined from the SDSS spectra, because the instrumental
resolution of $\sim$ 70 \kms\ sets a practical lower limit to the
\sigmastar\ that could be measured, and also because the $S/N$ of
the SDSS spectra was insufficient or the continuum was dominated by
AGN emission rather than by starlight.  Here we present new
measurements of \sigmastar\ for objects in the GH07 sample. These
new measurements provide a useful way to examine BH demographics,
even if the individual BH masses are not highly accurate. In
addition, our work provides a very large sample to examine
relationships between gas and stellar kinematics for nearby AGNs. We
present the sample properties, observations, and data reduction in
Section~\ref{sec:obsdata}, and measurements of \sigmastar\ and
line-width in Section~\ref{sec:meas}. We discuss the \msigma\
relation in Section~\ref{sec:mbhrel} and the narrow-line properties
Section~\ref{sec:nel}, with conclusions in Section~\ref{sec:con}.

\section{Observations and Data Reduction}
\label{sec:obsdata}

The objects observed at Keck and Magellan were selected from the
sample of low-mass BHs presented by \citet{GH07c}.  They selected
229 broad-line active galaxies with $\mbh <2\times\ 10^6 M_{\odot}$
from SDSS DR4. They estimated the BH masses with the single-epoch
virial method, using the Full Width Half Maximum (\fwhm) and
luminosity of the broad \hal\ emission line, following the methods
first described by \citet{GH05b}.  The virial calculation determines
the radius of the broad-line region (BLR) using the
radius-luminosity relationship of \citet{bentz06}.

The objects were observed during 2008--2009 using the Echellette
Spectrograph and Imager \citep[\esi;][]{she02} at the Keck-II
telescope and the Magellan Echellette \citep[\mage;][]{marsh08}
Spectrograph at the Magellan II Clay telescope at Las Campanas
Observatory. For each observation, the spectrograph slit was
oriented at the parallactic angle.  Flux standards and late-type
giant stars (F4--M0) for use as velocity templates were observed
during each night.  Details of the Keck and Magellan observations
and reductions are given below.  We obtained useful data for 65 of
66 objects observed with \mage\ and for 13 objects observed with
\esi; two objects were observed with both \esi\ and \mage.  We also
combine our new sample with the 17 similar objects previously
presented by \citet[][Hereafter BGH05]{barth05}.  Including the
BGH05 sample, our total sample consists of 93 low-mass SDSS AGNs.

\esi\ -- Observations with \esi\ at Keck were made during the nights
of 2008 March 1--2 UT.  We used a 0\farcs75 slit width, resulting in
an instrumental dispersion of $\sigma_{i} \approx 22$ \kms.  The
spectra cover the wavelength range 3800--10900 \AA\ across 10
echelle orders, and the dispersion is a constant 11.5 \kms\
pixel\per\ in velocity. The exposure times for individual objects
ranged from 900s to 3000s. One-dimensional spectra were extracted
within a 1\arcsec\ extraction width, and wavelength- and
flux-calibrated, with correction for telluric absorption bands,
using standard techniques following the same methods we have
previously used for ESI data (BGH05).

\mage\ -- Observations with \mage\ at the Magellan II Clay telescope
were carried out on the nights of 2008 April 9--13, 2008 August
28--30, and 2009 January 25--27 UT, using a 1\arcsec\ slit width,
giving an instrumental dispersion of $\sigma_{i} \approx 26$ \kms,
as measured from the arc lamp spectra.  The spectral coverage is
approximately 3200--10000 \AA\ across 15 echelle orders, with a
nearly constant dispersion of 23 \kms\ pixel\per.  Exposure times
ranged from 1800--7200 s and were typically 5400 s.  One-dimensional
spectra were extracted from the CCD images and wavelength-calibrated
using the \magerdc\ package kindly provided by George Becker.  The
extraction width is typically 1\farcs5 to 3\arcsec.  The \magerdc\
package uses techniques for rectification and sky subtraction
developed by \citet{kelson03}.  Optimal extractions \citep{Hor86}
were used for exposures of the bright stars, while for the galaxies
a simple boxcar extraction was used in order to avoid spurious
clipping of emission lines that can sometimes occur with optimal
extractions. Flux calibration and telluric absorption correction
were applied using the same methods used for the ESI data.  Finally,
cosmic rays were removed when combining multiple exposures.

\begin{figure}[h]
\begin{center}
\scalebox{0.35}{\includegraphics{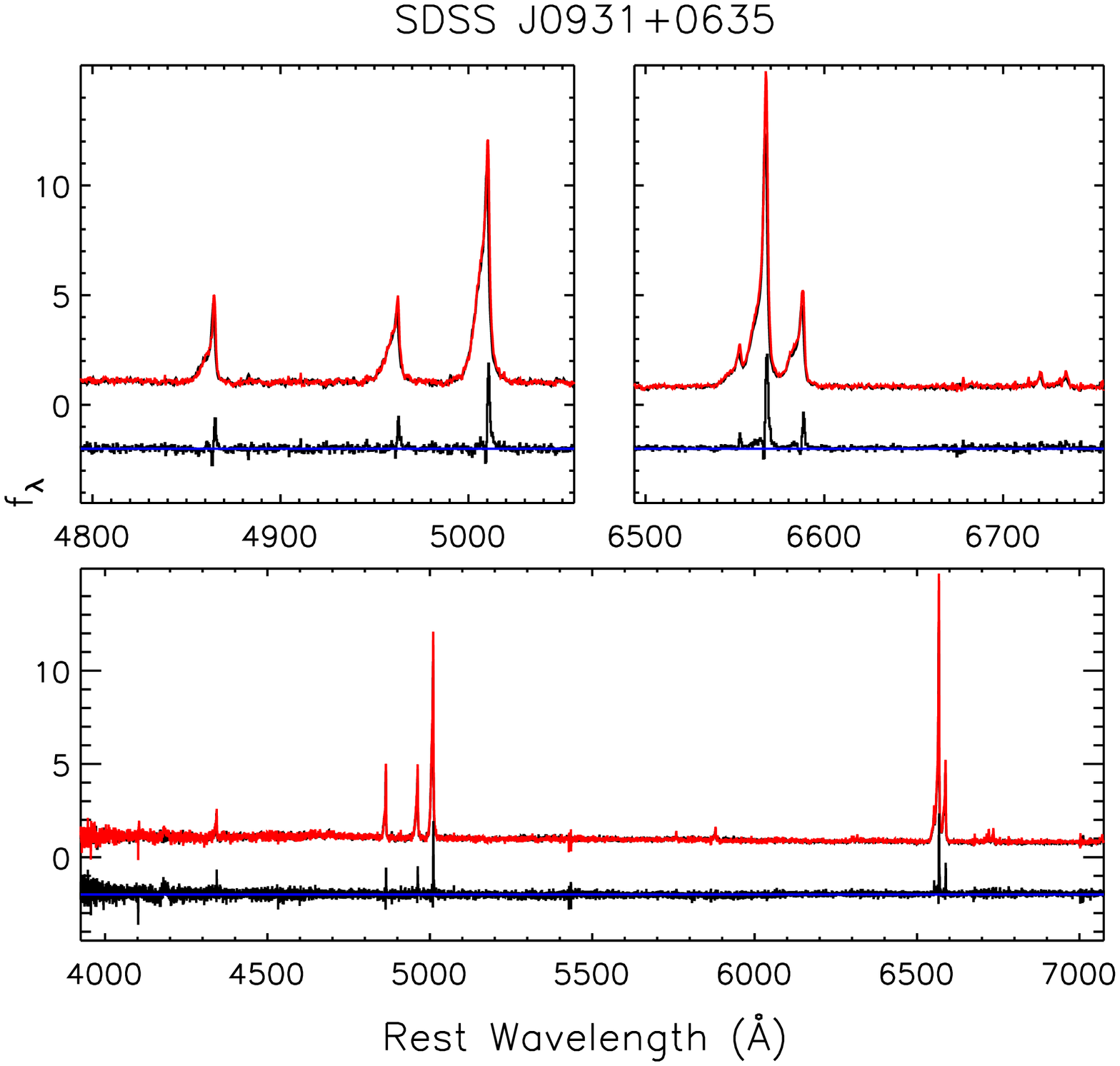}}
\scalebox{0.35}{\includegraphics{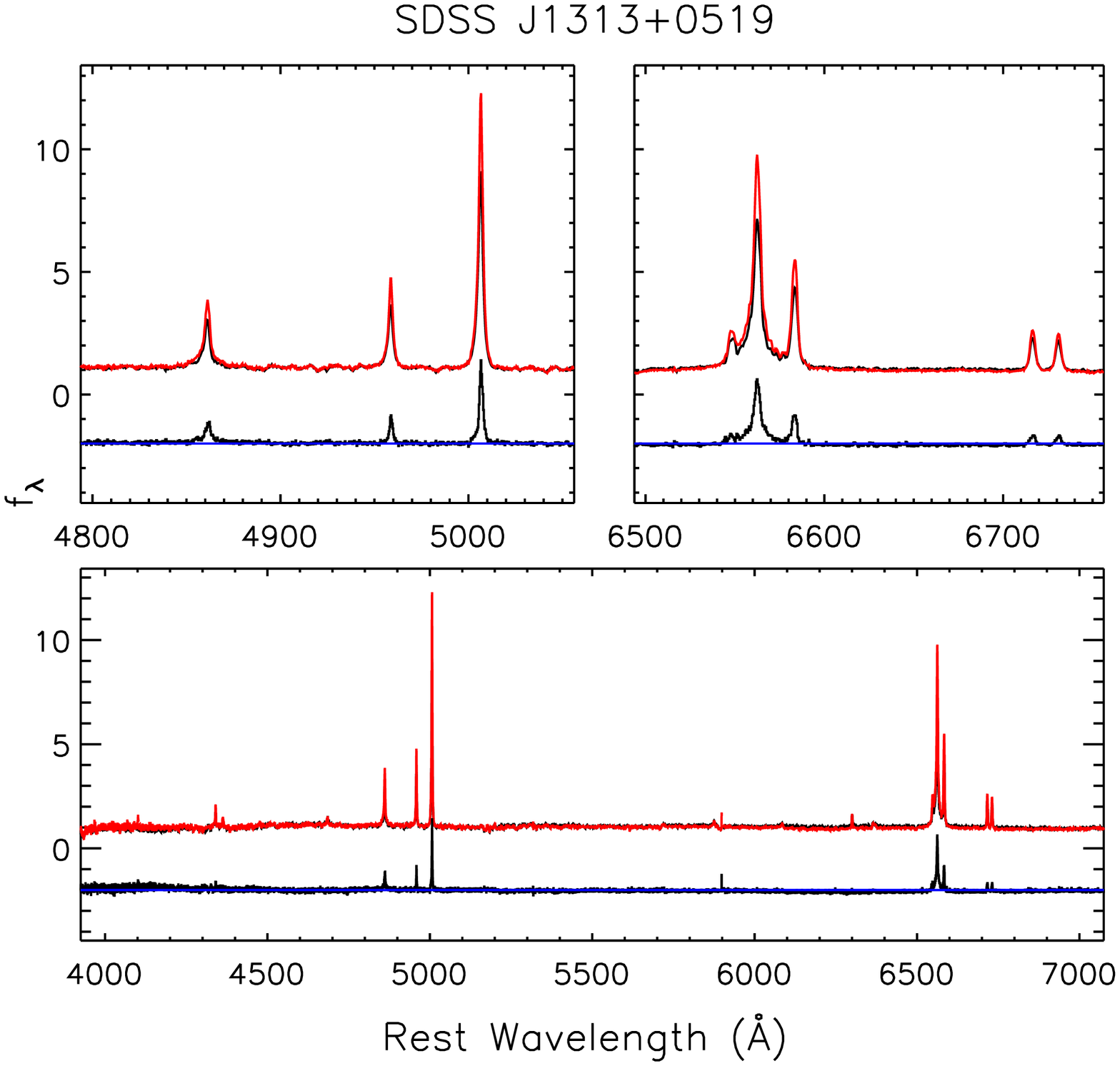}}
\end{center}
\caption{ Comparison of \mage\ and \esi\ spectra (in black and red,
  respectively) for SDSS J093147.25+063503.2 and SDSS
  J131310.12+051942.1.  The \esi\ data have been rebinned to the
  spectral resolution of the \mage\ data. To facilitate comparison, all
  spectra have been rescaled to have flux density of unity at 5100
  \AA. In each panel, the residual from subtraction of the \mage\
  spectrum from the \esi\ spectrum (minus an arbitrary constant for
  clarity) is shown as a black line in the bottom.  The blue horizontal line shows the arbitrary constant. } \label{fig:spec2}
\end{figure}

Two objects, SDSS J093147.25+063503.2 and SDSS J131310.12+051942.1,
were observed with both \esi\ and \mage, allowing a direct
comparison of the spectra; see Figure~\ref{fig:spec2}.  To
facilitate comparison, the \esi\ data have been re-binned to the
spectral resolution of the \mage\ data, and all the spectra have
been re-scaled to have flux density of unity at 5100 \AA.  The Keck
and Magellan spectra appear consistent across the entire available
wavelength range, which illustrates the consistency of the flux
calibration, except for differences in the emission lines, which are
expected due to the difference in instrumental dispersion of the two
instruments. For SDSS J093147.25+063503.2, the difference spectrum
reveals significant residuals only in the peaks of the narrow
emission lines, most likely resulting from the difference in
instrumental dispersion. For SDSS J131310.12+051942.1, there is also
evidence of some flux variation between the 2 observations.  This
could result from the different aperture size and different seeing
between the two exposures, but could also be in part due to real
variability of the continuum and broad emission lines.  However, the
time interval between these two observations is 329 days, and on
such a short timescale the narrow lines probably would not be
variable. Since there are residuals in the narrow lines in
Figure~\ref{fig:spec2} as well as the broad lines, the main cause is
probably aperture size and seeing.

\section{Measurements}
\label{sec:meas}

\subsection{Stellar Velocity Dispersions}
\label{sec:svd}

The stellar velocity dispersions were measured by a direct fitting
method \citep{burbidge61, rix92}, in which the spectra of velocity
template stars are broadened and fitted to the galaxy spectra
locally in a specific spectral region.  A Gaussian profile was
assumed as the line-of-sight velocity distribution.  We follow
\citet{barth02} and \citet{GH06a} and express the fitted model
spectrum, $M(x)$, as
\begin{equation}
M(x) = \{[T(x) \otimes\ G(x)] + C(x)\} P(x) ,
\label{eq:ctmmodel}
\end{equation}
where $T(x)$ is the stellar template spectrum, $G(x)$ is the
Gaussian broadening function, $C(x)$ represents a featureless
continuum, and $P(x)$ is a polynomial factor.  Since our fits are
performed across a small wavelength region, we adopt a quadratic
polynomial for $C(x)$, which could also account for some other
additive components, such as the ``pseudo-continuum'' due to
\ion{Fe}{2} emission originating from BLR of AGN.  The low-order
multiplicative polynomial $P$ allows for the differences between the
template and the galaxy, including continuum shape, reddening in the
galaxy spectrum, and wavelength-dependent flux calibration errors.
We use a quadratic polynomial for $P$, since higher-order
polynomials will tend to fit the absorption features and adversely
affect the dispersion measurements \citep{barth02}.  Besides the
velocity dispersion and six parameters for these two polynomials,
the redshift of the galaxy is also a free parameter.  We determined
the best-fit parameters by minimizing \chisq, using the
Levenberg-Marquardt least-squares fitting routine provided by the
\texttt{mpfit} package in IDL \citep{markwardt}.

Our collection of stellar templates includes 19 G and K giant stars
observed with \esi\ and 22 F, G, K and M giant stars observed with
\mage.  For each galaxy, we inspect the fitting and discard the
templates that fail to fit.  Then we list the \chisq\ of the
remaining templates in ascending order, and select the first
two-thirds to be well-fit templates.  The measured velocity
dispersion derived with the best-fitting template (minimum \chisq)
is adopted to be the best estimate of \sigmastar.  Generally, the
best fits were obtained with K-giant templates.  The uncertainty in
measurement is calculated as the quadrature sum of the fitting
uncertainty of the best-fit template and the standard deviation of
the measurements of all the selected templates.   A minimum of six
templates are used in each calculation, except for several objects,
marked by ``:'' behind the value of \sigmastar\ in
Table~\ref{table1}, that were not well fit by six or more templates.
In these cases, all the results of well-fit templates will be
accounted in estimating the uncertainty.

The spectral region around the \ion{Ca}{2} $\lambda\lambda8498,
8542, 8662$ triplet ($\sim8470-8700$ \AA, hereafter the CaT region)
is ideal for measuring velocity dispersions because the CaT lines
are strong, not blended with other strong lines, and relatively
insensitive to stellar population variations.  They are also less
strongly diluted by AGN continuum contamination than stellar
features at blue wavelengths. However, for objects with redshift
higher than 0.05, the CaT absorption features can be affected by
both night sky emission residuals and telluric absorption bands,
making it difficult to obtain useful measurements.
Stellar features at bluer wavelengths are unaffected by telluric
absorption, and we also carried out measurements in the \mgb\ region
($\sim5050-5430$ \AA) and the ``Fe region'' ($\sim5250-5820$ \AA)
redward of \mgb\ for measurements, following \citet{barth02} and
\citet{GH06a}.  \citet{GH06a} ran a series of simulations to
evaluate the contamination of narrow emission lines, including
[\ion{Fe}{7}] $\lambda5158$ \AA, [\ion{Fe}{6}] $\lambda5176$ \AA, and
[\ion{N}{1}] $\lambda\lambda5197,5200$ \AA\AA, around the \mgb\ features,
as well as the pseudo-continuum of broad \ion{Fe}{2} extending in
this region. They found that the Fe region containing strong
\ion{Fe}{1} absorption features resulted in better recovery of
\sigmastar, since it includes less contamination by coronal emission
lines.

We tested these two regions by fitting stellar templates to a set of
simulated spectra, composed as a linear combination of a K2 star, an
A0 star, and a featureless linear continuum.  The combined spectra
were then broadened by a Gaussian velocity-broadening kernel with
width ranging from $\sigmastar = 30$ to 100 \kms\ in increments of
10 \kms.  Slight variations in the shape of the spectra,
representing calibration errors and random errors, were imposed on
the broadened model spectra, and the models were degraded to $S/N =
10$, 30, 50, and 100 per pixel.  We also redshifted the model
spectra by an arbitrary value comparable to the redshifts our
observed sample. Then the modeled spectra were fitted using the
methods described above, and the measured \sigmastar\ was compared
with the input value. We find, for $\delta\sigmastar \equiv
[\sigma(\textrm{output})-\sigma(\textrm{input})]/\sigma(\textrm{input})$,
that $\langle\delta\sigmastar\rangle = 0.02\pm0.11$ from the \mgb\
region and $\langle\delta\sigmastar\rangle = 0.03\pm0.02$ from the
Fe region for $S/N = 30$, the median $S/N$ for our observed sample.
Reassuringly, the results from these test measurements are
consistent with input values.  We find that the scatter in results
from the \mgb\ region is larger than that from Fe region.  The
scatter for both fitting regions decreases with increasing $S/N$, to
0.04 and 0.01 for $S/N=100$, respectively.  The larger scatter of
the \mgb\ region might be attributed to template mismatch.  We
checked the template spectra and found that for different types of
stars, the variations in the width of the \mgb\ lines are slightly
larger than the width variations for the \ion{Fe}{1} features.  For
different type of stars, for example, in the sequence of G8
$\rightarrow$ K3 $\rightarrow$ K5, the \mgb\ line widths increase,
while Fe absorption widths are consistent; this may explain at least
in part why the dispersion measurements obtained from the \mgb\
region have larger scatter.

In practice, when fitting to actual galaxy spectra, we excluded
wavelength regions covering emission lines including [\ion{N}{1}]
$\lambda\lambda5197,5200$ \AA\AA\ and the high-ionization Fe lines.  The
majority of the objects have several high-ionization Fe lines,
including [\ion{Fe}{7}] $\lambda5158$ \AA, [\ion{Fe}{6}] $\lambda5176$ \AA,
[\ion{Fe}{7}] $\lambda5278$ \AA, [\ion{Fe}{14}] $\lambda5303$ \AA\ and
[\ion{Fe}{6}] $\lambda5335$ \AA.  We tested model fits that excluded and
included the \mgb\ absorption features themselves, and found that
the best-fit template could generally fit the \mgb\ features well
for galaxies not dominated by AGN Fe emission.  This is because most
objects in this sample have relatively low \sigmastar, so the
mismatch of [Mg/Fe] abundance ratio that sometimes affects template
fits to elliptical galaxies with high \sigmastar\ \citep{worthey} is
apparently not a significant issue for this sample.

\begin{figure*}[h]
\begin{center}
\scalebox{0.58}{\includegraphics{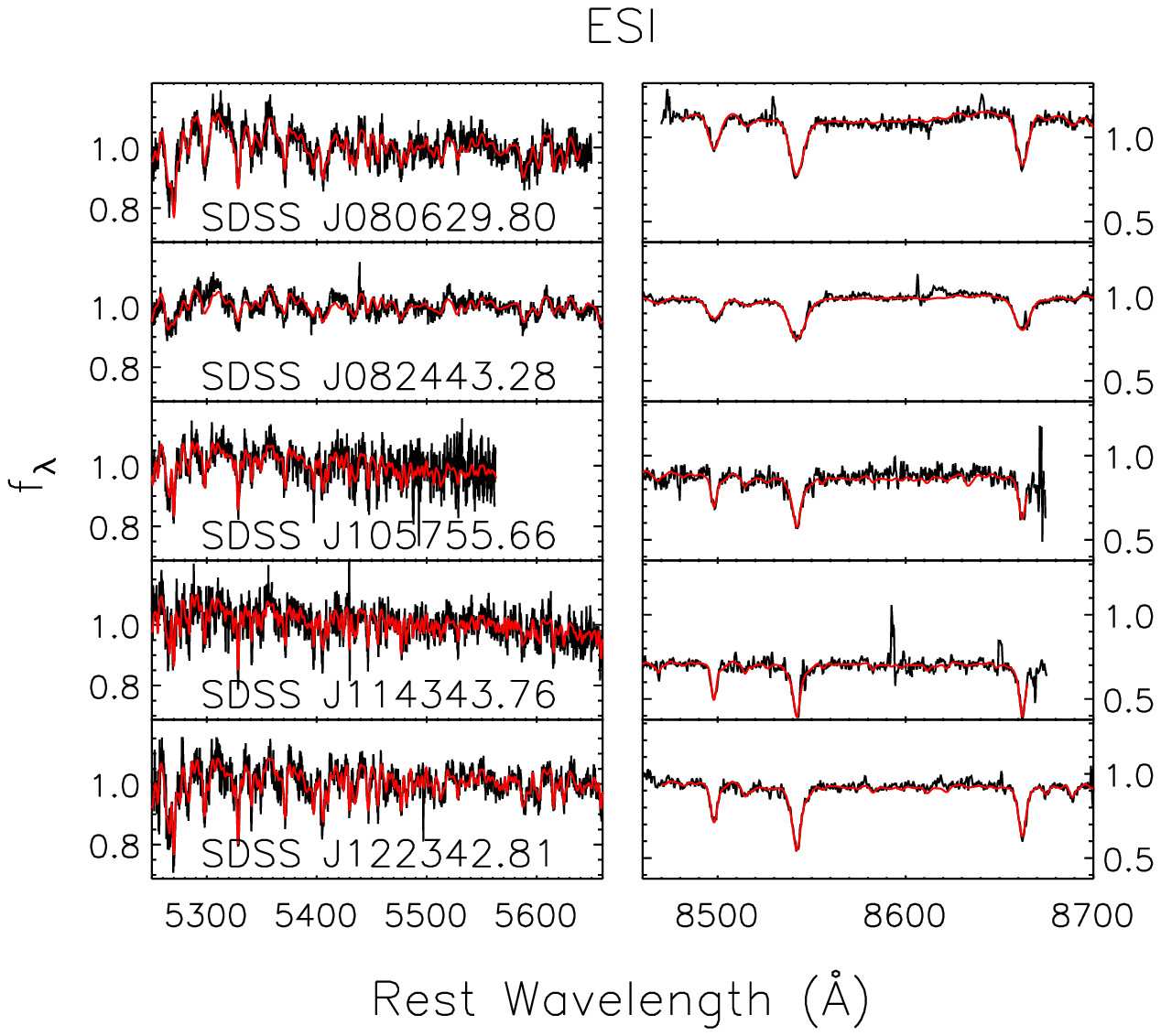}}
\scalebox{0.58}{\includegraphics{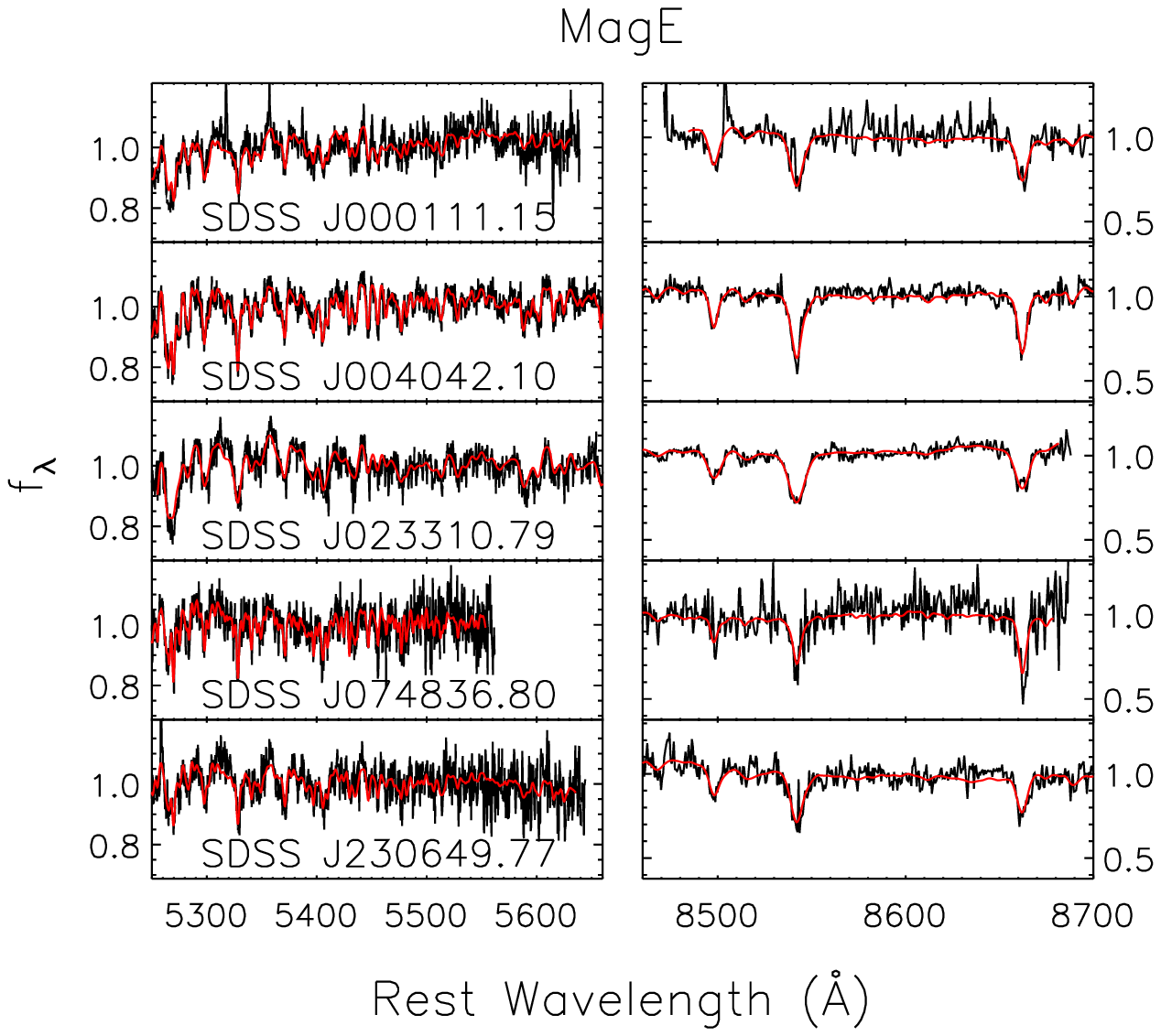}}
\end{center}
\caption{ Examples of \esi\ (left) and \mage\ (right) spectra of
  spectra in the Fe and CaT regions. The spectra are flux-calibrated
  in $f_{\lambda}$ units and normalized to a flux level of unity.  The
  observed spectrum is in black and the best-fitting broadened stellar
  template is in red.
   } \label{fig:vdisp}
\end{figure*}

In order to avoid issues of mismatch in flux calibration or spectral
resolution across echelle orders, we prefer to measure velocity
dispersions from a single echelle order (rather than from multiple
orders that have been ``stitched'' together), and the specific fitting
region for each individual galaxy was adjusted to remain within one
echelle order. For \mage\ data, we first measured \sigmastar\ from
both the \mgb\ region and the Fe region individually.  We found that
if these two regions are in the same order, we obtained highly
consistent results from both regions.  This is partly due to the
limited wavelength range of each order for \mage\ data and the fact
that there is substantial overlap between the \mgb\ and Fe regions.
Therefore, if these two are on the same order for a given galaxy, we
carry out a single fit extending across both of these regions, and
list the result as \sigmastar(\mgb) in Table~\ref{table1}.  When a
separate value is listed in the \sigmastar(Fe) column of the table,
this denotes that the \mgb\ and Fe regions fell in adjacent echelle
orders and were fitted separately.  Figure~\ref{fig:vdisp} illustrates
some examples of both \esi\ and \mage\ spectra in the Fe and CaT
regions.

We were able to obtain useful measurements of \sigmastar\ for 56 of
the 76 newly observed galaxies.  The two galaxies with both \esi\
and \mage\ observations have consistent velocity dispersions within
the uncertainty measured by two instruments.  The remaining 20
galaxies are either dominated by AGN emission or have $S/N$ too low
to permit a successful fit.  We also re-measured \sigmastar\ for 15
of the 17 objects from BGH05 (excluding the two objects from that
sample that were highly AGN-dominated). The results are consistent
with the BGH05 measurements, within the uncertainties.  If we define
the deviation between our new measurements and BGH05 both from the
\mgb\ region, as $\Delta \sigmastar = \log \sigmastar(\mathrm{new})
- \log \sigmastar(\mathrm{BGH05})$, the mean value of $\Delta
\sigmastar$ is less than 0.001 dex and the rms difference is only
0.02 dex.  All of the \sigmastar\ measurements are summarized in
Table~\ref{table1}. For the galaxies with more than one measurement
of \sigmastar\ from different fitting regions, the corresponding
results are mostly consistent within the uncertainties. We take the
average of the available measurements as the best estimate of the
stellar velocity dispersion in each object, and list as \sigmastar\
in Table~\ref{table1}.

\subsection{Emission-line Properties}
\label{sec:emline}

To decompose the \hn\ lines, we first subtracted the
underlying continuum.  The continuum model is essentially the same
as that in Equation~\ref{eq:ctmmodel} used to measure \sigmastar,
but fitted over the spectral region surrounding \hal, covering the
rest-frame region 6100--7100 \AA.  For this model spectrum ($M(x)$
in Equation~\ref{eq:ctmmodel}), the velocity dispersion was
constrained to lie within three times the $1\sigma$ uncertainties
around the measured velocity dispersion as described in the
preceding section.  Emission lines were masked out from the
calculation of $\chi^2$ for the fits.  The continuum fits were
carried out on spectra in which the echelle orders were ``stitched''
together, since in some cases the \hal\ line falls near the end of
an echelle order.  The best-fit continuum model was then subtracted
from the spectra to yield a pure emission-line spectrum.

We used multiple-Gaussian models to fit the \hal\ and [\ion{N}{2}]
lines as well as [\ion{S}{2}] $\lambda\lambda6717,6731$ \AA\AA.  We
followed \citet{GH07b}, using a multi-component Gaussian fit to the
[\ion{S}{2}] doublet to model other blended lines.  Up to 4
Gaussians were used to model the [\ion{S}{2}] doublet.  The
velocities and widths of the two [\ion{S}{2}] lines were constrained
to be the same, while the intensity ratio was allowed to vary.  For
objects with very weak or absent [\ion{S}{2}] emission, we used the
core of the [\ion{O}{3}] $\lambda5007$ \AA\ line as the narrow-line model
instead.

\begin{figure}[h]
\begin{center}
\scalebox{0.45}{\includegraphics{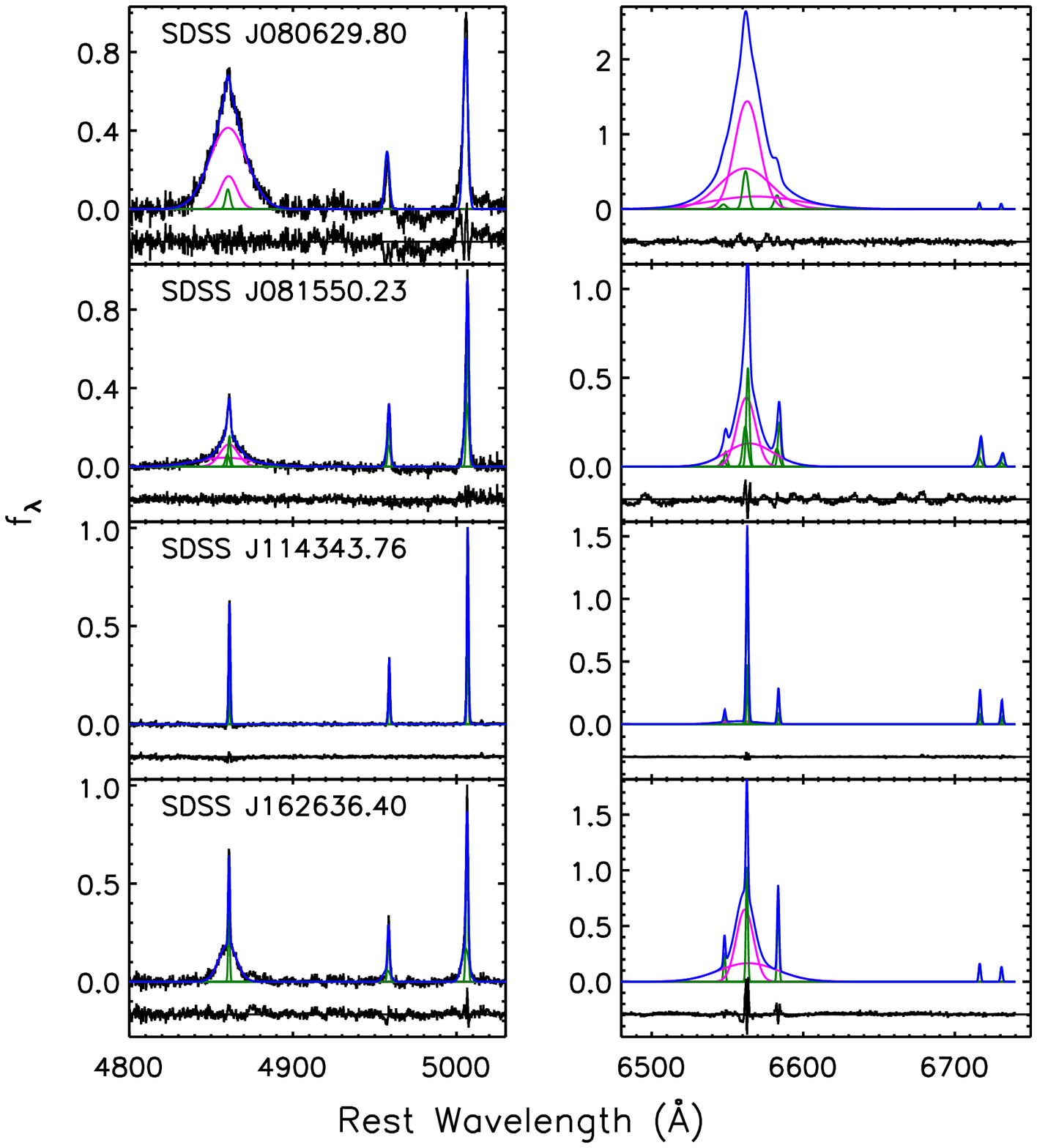}}
\scalebox{0.45}{\includegraphics{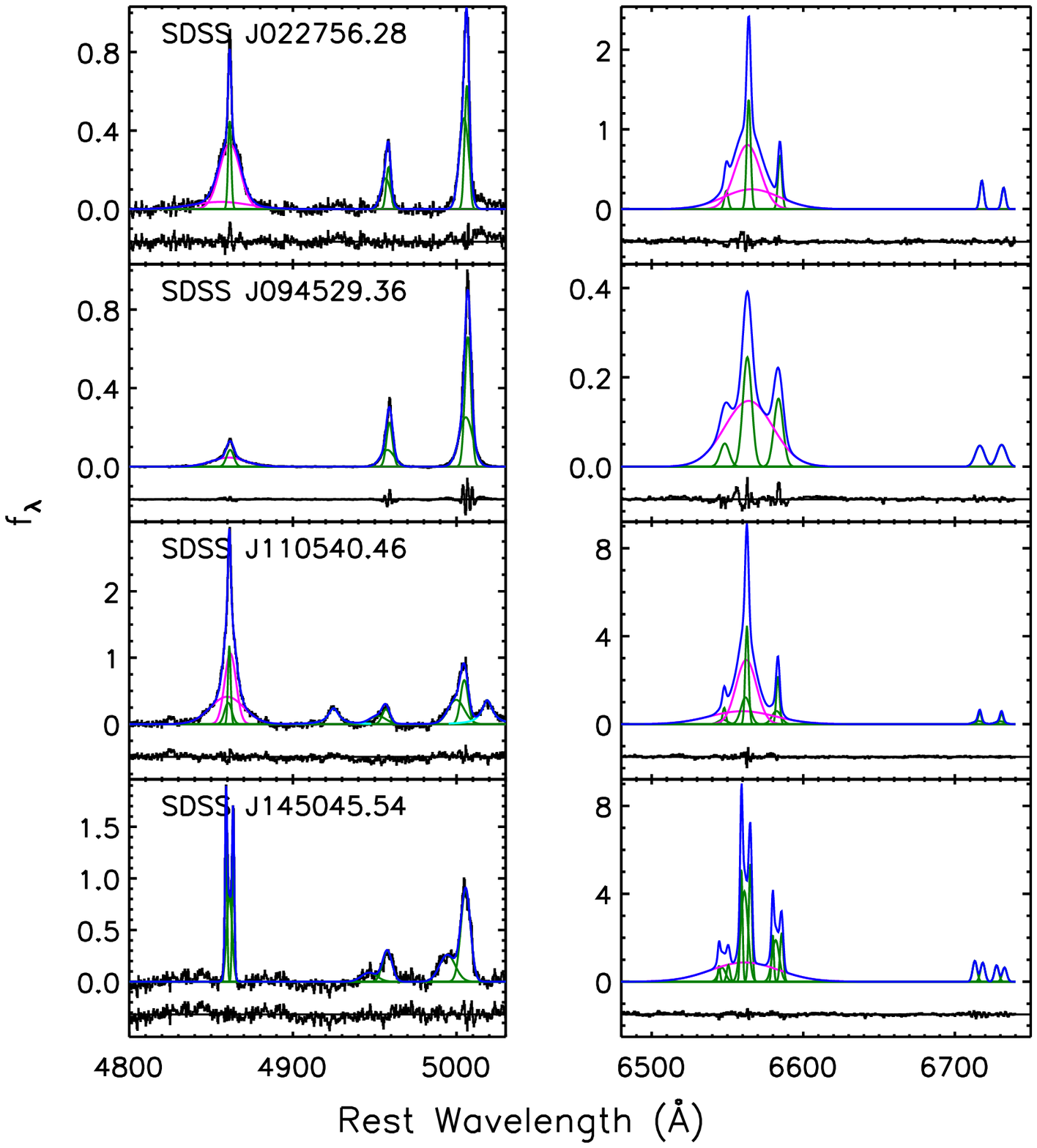}}
\end{center}
\caption{ Examples of fitting to the \hbe\ + [\ion{O}{3}] and \hn\ regions for \esi\ (left) and \mage\ (right) data.
  The spectra are flux-calibrated in $f_{\lambda}$ units and arbitrarily
  normalized to a flux level of unit around the peak of [\ion{O}{3}] $\lambda5007$ \AA.  In
  each panel, continuum-subtracted spectra (black) and the best-fit
  model (blue) are shown, with their residuals shifted downward by an
  arbitrary constant for clarity.  Individual components are
  overplotted: narrow lines (green), broad lines (magenta) and
  \ion{Fe}{2} emission (cyan) if present. } \label{fig:linefit}
\end{figure}

The [\ion{N}{2}] doublet lines were constrained to have their relative
rest-frame wavelengths and intensity ratio fixed to their laboratory
values of 35.42 \AA\ and 2.96, respectively.  We then fit as many
Gaussian components to the broad component of \hal\ as needed to
achieve an acceptable fit: starting with a single Gaussian model, new
components were added one at a time if they resulted in 20\% decrease
in \chisq. Generally, one or two Gaussians proved sufficient to fit
the broad \hal\ emission adequately. Figure~\ref{fig:linefit} shows
some examples of fits to \hal.  Only a small number of galaxies with
asymmetric profiles or very broad wings required a third Gaussian
component.  In most cases we were able to achieve an acceptable fit to
the \hn\ blend, but there are a few cases ($\sim6$) for
which no acceptable fit to the narrow \hal\ line could be obtained
with the [\ion{S}{2}] model.  For those cases, we relaxed the width
constrains on the narrow lines.  In each case, the width of the narrow
\hal\ in the best-fit model remained quite close to the [\ion{S}{2}]
line width but the \hal\ fit was significantly improved by allowing
its width to be a free parameter.

We also modeled the \hbe\ + [\ion{O}{3}] region in
continuum-subtracted spectra, in the rest-frame range of 4800-5150
\AA, following procedures similar to those described above for the
\hn\ region.  We used double-Gaussian models for the
broad component of \hbe\ as well as for [\ion{O}{3}], for which the double Gaussians represent
a wing and a core component. The corresponding components in the two
[\ion{O}{3}] lines are constrained to have the same velocity width,
and the wavelength separations and intensity ratios are fixed at
their laboratory values.  For those galaxies with significant broad
\ion{Fe}{2} emission (mainly \ion{Fe}{2} $\lambda\lambda4924,
5018$ \AA\AA), we adopted an analytical model for the broad Lorentzian
system ``L1'' in \citet[][see their Appendix A]{vv04}.  Each
\ion{Fe}{2} line was fitted with a Lorentzian profile, allowing the
line center to vary within a small range around the expected value
to account for shifting from the systemic redshift.  The widths of
both \ion{Fe}{2} lines were constrained to be the same, while their
flux ratio was allowed to vary.  In order to model the narrow \hbe,
we follow \citet{GH05b} to use the profile of [\ion{S}{2}].  The
\hbe\ centroid was fixed to the relative wavelength of narrow \hal\
if necessary, and its flux was limited to be no larger than the
value for Case B recombination \citep[\hal\ = 3.1\hbe,][]{oster89}.
The procedure produced acceptable fit to \hbe\ in most cases, except
for some objects ($\sim13$) in which the profile of [\ion{S}{2}]
seems not be a good model for narrow \hbe, possibly resulting from
small changes in spectral resolution across different echelle orders
of the spectra. In these cases we used a single Gaussian to model
narrow \hbe, and we relaxed the width and flux constraints.  In
Figure~\ref{fig:linefit} we show some examples of best-fit models
for the \hbe\ + [\ion{O}{3}] region.  Individual components for
\ion{Fe}{2} emission are also shown if they are present.  It turns
out that the detected \ion{Fe}{2} lines in this sample are
relatively narrow, typically 700 \kms\ \fwhm, significantly narrower
than the commonly used \ion{Fe}{2} template I ZW 1, whose broad line
system ``L1'' has $\fwhm = 1100$ \kms\ \citep{vv04}.

The \fwhm\ of the overall profile of broad \hal\ is used for tracing
the velocity dispersion of gas in the BLR, in the close environment
of the black hole.  We continue to use the \fwhm(\hal) as a measurement
of the line width for the whole sample in a consistent way, because the
commonly used \hbe\ lines suffer from low $S/N$.  Broad \hbe\ is
often weak and sometimes not even detectable for objects in our
sample.

To estimate errors on the line width measurements, we should include
statistical uncertainties from profile fits caused by noise and
other random errors from spectral extraction and calibration, and
uncertainties in continuum subtraction and deblending of broad
components from narrow emission lines.  In general, the formal
fitting uncertainties from the profile fits are quite small,
typically only 11 \kms.  Since the broad lines for this low-mass
sample are narrow (mostly with \fwhm\ $<$ 2000 \kms), the
measurements of widths are much less affected by continuum
subtraction than broad-lined AGN \citep[see Section~2.5 of
][]{dong08}.  Due to the small wavelength range over which the
continuum-subtraction fits were performed and the good quality of
the fits, the continuum subtraction does not add substantially to
the error budget for the line widths.  To explore the uncertainty
from the emission-line profile deblending in more detail, we create
a set of artificial spectra from the best-fit model, following
\citet{GH05b}.  For each galaxy, we created a realization of the
combined emission lines from the best-fit model parameters, and
added Gaussian noise to match the $S/N$ of the data.  The artificial
spectra created this way suffer from the same deblending difficulty
as the original data.  Then the artificial spectra are fitted with
multiple-Gaussian models using the fitting procedure described
above.  The difference between the model and measured \fwhm(\hal) is
typically $\sim\ 5\%$ (or 0.02 dex).  With these artificial spectra,
we also investigated the uncertainty associated with the choice of
model for the narrow-line profile, which often dominates the
uncertainty in the decomposition of narrow/broad lines. We
substituted the narrow-line model with the [\ion{S}{2}] profile, the
core of [\ion{O}{3}], and narrow \hbe\, respectively, in the fitting
procedure with the multiple-Gaussian model above.  The typical
standard deviation in the \fwhm\ of broad \hal\ is $\sim\ 8\%$ (or
0.03 dex).  These two sources of error are combined to be typically
$\sim\ 9\%$ (or 0.04 dex), and taken as the estimated uncertainty of
\fwhm(\hal).

\subsection{Comparison of Broad-Line Widths With GH07}

It is instructive to compare our broad \hal\ measurements with those
obtained from SDSS spectra by GH07.  The smaller spectroscopic
apertures for the Keck and Magellan data result in a smaller degree of
starlight dilution of the AGN features, and the higher spectral
resolution of the new data should permit more accurate deblending of
the emission lines.  Thus, we expect that the Keck and Magellan data
should generally yield more accurate measurements of the broad \hal\
widths, particularly for very weak \hal\ emission lines.  The SDSS
spectra, on the other hand, have a more reliable flux calibration.  In
a small fraction of the objects from the GH07 sample, the broad \hal\
emission in the SDSS spectra is so weak that it is uncertain whether a
distinct broad component is genuinely present or not, and our new
spectra are particularly useful for testing the reality of these
features tentatively seen in the SDSS data.

\begin{figure}[h]
\begin{center}
\scalebox{0.6}{\includegraphics{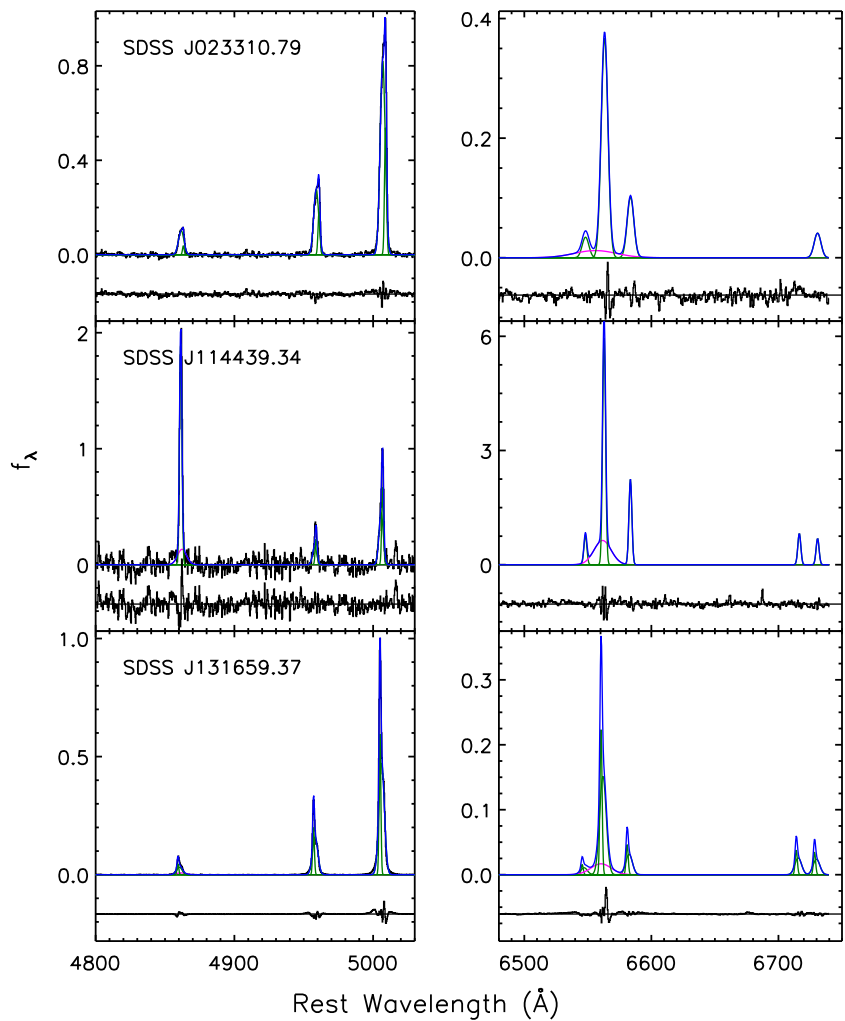}}
\end{center}
\caption{ Examples of galaxies showing possible broad \hal\
  components. Fit components are illustrated as in
  Figure~\ref{fig:linefit}.} \label{fig:broadha}
\end{figure}

We examined how our new measurements of the broad \hal\ \fwhm\
compare with the results that GH07 obtained from fitting SDSS
spectra, to test whether the objects with very weak broad \hal\ in
GH07's low-mass AGN sample were genuine Seyfert 1 galaxies or not.
We visually inspected the best-fit models for all the objects and
divided them into two categories: objects with ``definite'' and
``possible'' broad \hal.  In the following we refer to the two
sub-categories as the $d$ and $p$ sub-samples, as indicated in
Table~\ref{table1}.  For the $d$ sample, the wings of broad \hal\
generally extend beyond the [\ion{N}{2}] emission lines, and the
broad \hal\ is significantly wider than the narrow emission lines
such as [\ion{N}{2}].  We classified objects as belonging to the $p$
subsample if the peak amplitude of broad \hal\ was less than twice
the RMS pixel-to-pixel deviation in the continuum-subtracted
spectrum in the region surrounding \hal, or if the ratio of the flux
of broad \hal\ to the RMS deviation of the continuum-subtracted
spectrum was below 200.  Some examples of ``possible'' broad \hal\
are shown in Figure~\ref{fig:broadha}.

Among the 93 galaxies in our sample, 14 (15\%, 1 galaxy observed
both by \esi\ and \mage) had ambiguous ``possible'' broad \hal\
emission.  To investigate whether this was simply due to these 15
spectra having low $S/N$, we examined the average $S/N$ in the \hal
+ [\ion{N}{2}] region, and found that only three $p$ objects had very low
$S/N$ of $<11$.  Overall, the objects classified as being in the $p$
category have a median $S/N$ of 25, compared to a median value of 40
for our whole sample, so some of the $p$ objects may simply be
suffering from low $S/N$.  However, some of the $p$ objects have
very high $S/N$ spectra, so we can not attribute all the cases of
ambiguity in broad \hal\ to low $S/N$. There are six $p$ objects which
only show very weak or possible broad \hal\ in the SDSS spectra
(these are denoted as the $c$ subsample in GH07).  The other 8 out
of the 14 $p$ galaxies are classified as broad-line AGNs by GH07.
The difference between the SDSS results and our new fitting results
could possibly be due to intrinsic AGN variability or variable AGN
obscuration. Among the eight $p$ galaxies, SDSS J093147.25+063503.2 is
confirmed to have outflow components in the narrow lines, which
might have mimicked part of the broad component in SDSS spectra.
There are also five galaxies in the $c$ (``candidate'') sample of GH07
which we now classify as having definite broad \hal\ based on our
new high-resolution spectra.  A likely explanation is that the
better quality of our new spectra makes the decompositions more
accurate and better reveals the intrinsic properties of the broad
components.  Therefore, eight (9\%) out of 82 galaxies previously
classified as broad-line AGNs by GH07 are found to have only
possible broad \hal\ from the high-resolution spectra.  This
indicates the likely rate of false positive detections of broad
\hal\ in the SDSS sample for objects near the threshold of
detectability for broad \hal.  In our sample, there is one object,
SDSS J145045.54$-$014752.8, showing obvious double-peaked features
in narrow lines, and also a definite broad component in \hal\ (see
the last object of the right panel in Figure~\ref{fig:linefit}).

\begin{figure}[h]
\begin{center}
\scalebox{0.6}{\includegraphics{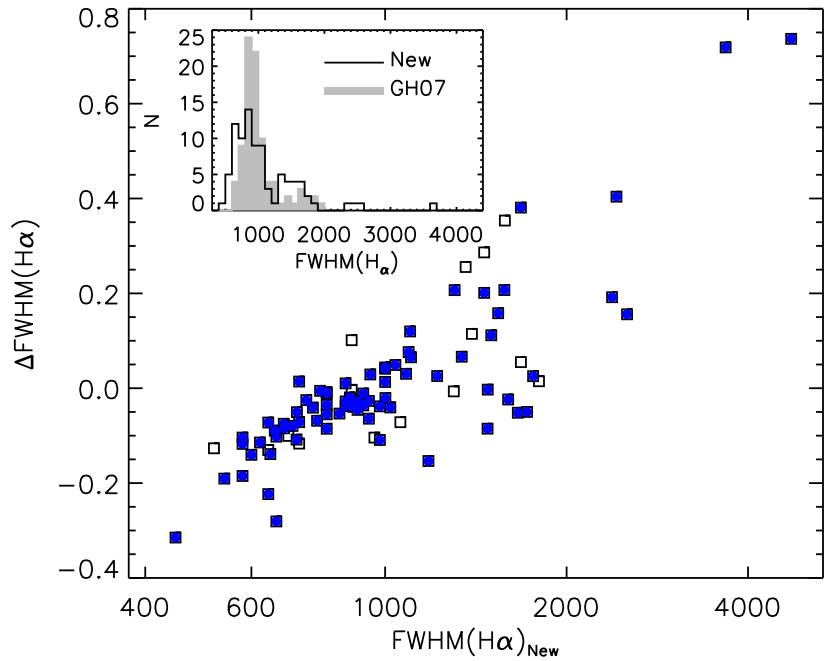}}
\end{center}
\caption{ $\Delta \fwhm$(\hal), the logarithmic difference between
  the GH07 measurement of broad \hal\ \fwhm\ and our new measurement,
  as a function of broad \hal\ linewidth.  Filled and open squares
  represent galaxies with definite and possible broad \hal,
  respectively.  The inset box shows the distributions of broad \hal\
  linewidths from GH07 and from our new measurements. }
\label{fig:fwhmha}
\end{figure}

We compare our new measurements of broad \hal\ FHWM to those of GH07
by defining $\Delta \fwhm \equiv \log \fwhm$(\hal)$_\mathrm{new} -
\log \fwhm$(\hal)$_\mathrm{GH07}$.  Figure~\ref{fig:fwhmha} displays
$\Delta$\fwhm\ vs. \fwhm(\hal)$_\mathrm{new}$.  These two measurements
are in reasonable agreement up to \fwhm\ $\thicksim$ 1000 \kms, but
the difference increases as \fwhm\ increases beyond 1000 \kms.  This
is probably because the higher resolution of the new data enables us
to fit the emission lines more accurately with more complex
multi-Gaussian models, although some of the difference may be due to
intrinsic source variability as well.

\section{Black Hole Masses and the \msigma\ Relation}
\label{sec:mbhrel}

In this section we discuss our method of estimating black hole
masses, and investigate the \msigma relation for our sample.  In the
following formulae for $M_{BH}$ estimation, \lctm\ denotes the AGN
continuum luminosity $\lambda$L$_{\lambda}$ at $\lambda = 5100$~\AA.

\subsection{Black Hole Mass}
\label{sec:mbh}

The black hole mass is estimated via the virial relationship, $\mbh\
= f\rblr\Delta V^2 /G$, where \rblr\ is the radius of the BLR, and
the orbital velocity at that radius is estimated by the velocity
width of the broad emission line, $\Delta V$.  Assuming that the gas
in the BLR is virialized, the gas velocity traces the central mass
in AGN, which is dominated by the black hole within the BLR radius.

The broad line width can be measured with the \fwhm\ or $\sigma_{\rm line}$,
which is the second moment of the profile, or the line dispersion.
Both have merits and difficulties \citep{peterson04}.  The line
dispersion $\sigma_{\rm line}$ has been
suggested to be a more robust and precise estimator of viral
velocity when measured from the RMS spectra of reverberation mapping
datasets \citep{peterson04, collin06}.  However, it is highly sensitive
to the contribution from the extended line wings, and therefore less
robust in single-epoch spectra when there is blending of other
emission lines on the line wings \citep{denney09}.  For measurements
from single-epoch data, the \fwhm\ is more commonly adopted \citep[see a review in][]{mcgill08}; it is sensitive to the line core and to the
decomposition of the broad and narrow components, but it is
relatively insensitive to the accuracy of measurement of faint
extended wings on the broad-line profile.  Denney \etal\ suggests
that when the $S/N$ is lower than $10-20$, both line width
measurements would become unreliable, and
line-profile fits would introduce systematic errors to single-epoch masses
(e.g., $\sim0.17$ dex offset in \mbh\ estimated with \fwhm, their
Table 5).  In practice, we generally fit the broad \hal\ with
two Gaussians. This is a purely empirical procedure and no specific
physical meaning is assigned to the two components separately. Thus,
we measure \fwhm(\hal) from the overall profile of the Gaussian
components used to fit the broad component of \hal.

We follow GH07's approach \citep[see][Appendix]{GH07c} to calculate
\mbh, which is estimated by using the line-width of \hal\ and the BLR
radius inferred from the broad \hal\ luminosity \citep{GH05b}.  In
this work, the width of the broad \hal\ emission line is measured by
decomposing the \hn\ lines in the \esi\ and \mage\
data.  The method relies on the broad-line region radius-luminosity
(\rl) relation derived from reverberation mapping \citep{kaspi05,
bentz06, bentz09} to determine the BLR radius from the estimated
continuum luminosity, which in turn is estimated from the broad
\hal\ luminosity following \citet{GH07c} since this is more
accurately determined than the nonstellar continuum from our
spectroscopic data. We update GH07's ``recipe'' for determining BH
masses with the revised \rl\ relation presented by \citet{bentz09},
who analyzed high-resolution \hst\ images of 34 reverberation-mapped
(RM) AGNs to obtain more accurate measurements of AGN continuum
luminosity.  The revised \rl\ relation from \citet{bentz09} is
\begin{equation}
\log \left( \frac{\rblr}{\textrm{lt-days}} \right) =
1.50^{+0.05}_{-0.02} + 0.519^{+0.063}_{-0.066} \log \left(
\frac{\lctm}{10^{44}~\textrm{\ergs}} \right),
\end{equation}

We follow GH07 in assuming $f=0.75$ \citep{Netzer90} although we
note that this $f$ value is not derived for any specific physical
model of the BLR.  The choice of this particular $f$ factor aids in
comparison of our results with prior work, as described below.  From
\citet{GH05b}, the empirical relation between the line widths of
broad \hal\ and \hbe\ is

\begin{equation}
\fwhm(\textrm{\hbe}) = (1.07 \pm 0.07) \times 10^3 \left(
\frac{\fwhm(\textrm{\hal})}{10^3 ~\textrm{\kms}} \right)^{1.03 \pm
0.03} ~\textrm{\kms} .
\end{equation}

Combining these results with the virial relationship gives the BH mass as

\begin{equation}
\log \left( \frac{\mbh}{\msun} \right) = 6.72^{+0.08}_{-0.06} +
0.519^{+0.063}_{-0.066} \log \left( \frac{\lctm}{10^{44}
~\textrm{\ergs}} \right)  + (2.06 \pm 0.06) \log \left(
\frac{\fwhm(\textrm{\hal})}{10^3~\textrm{\kms}} \right) .
\label{eq:mbh-lctm}
\end{equation}

If we substitute the continuum luminosity with luminosity of \hal\
using the following empirical relation from \citet{GH05b},

\begin{equation}
\lhal = (5.25 \pm 0.02) \times 10^{42} \left( \frac{\lctm}{10^{44}
~\textrm{\ergs}} \right) ^ {1.157 \pm 0.005} ~\textrm{\ergs} ,
\label{eq:lhalctm}
\end{equation}

then we obtain the BH mass as

\begin{equation}
\log \left( \frac{\mbh}{\msun} \right) = 6.40^{+0.09}_{-0.07} +
(0.45 \pm 0.05) \log \left( \frac{\lhal}{10^{42} ~\textrm{\ergs}}
\right) + (2.06 \pm 0.06) \log \left(
\frac{\fwhm(\mathrm{H}\alpha)}{10^3 ~\textrm{\kms}} \right)   .
\label{eq:mbh}
\end{equation}

Our Keck and Magellan spectra were not all taken under photometric
conditions, and since the observations were obtained through narrow
spectroscopic apertures, slit losses can be significant.  The SDSS
spectra have a more consistent flux calibration, so it is preferable
to use $L$(\hal) measured from the SDSS data, even if this does
introduce some additional uncertainty due to the fact that the \hal\
linewidths and luminosities are measured from non-simultaneous
observations.  Most objects in our sample are in the GH07 sample of
active galaxies containing low-mass BHs, so we can obtain \lhal\ for
most of our sample from GH07.  For the 5 objects in BGH05 sample
that were not included in GH07, we use the \hal\ luminosity \lhal\
from \citet{GH04}.  One object in the BGH05 sample was not part of
either the GH07 or \citet{GH04} catalogs, and for this object we
estimated \lhal\ from the \lctm\ measured by \citet{barth05} using
the \lhal-\lctm\ relation (Equation~\ref{eq:lhalctm}) of
\citet{GH05b}.

The BH masses estimated by Equation~\ref{eq:mbh} are systematically
lower by 0.08 dex than those obtained using the method described in
the Appendix of \citet{GH07c}, since they used the earlier version
of the \rl\ relationship from \citet{bentz06}.  We still refer to
our mass estimator as $M_{\mathrm{GH07}}$, since it follows their
basic method with only the \rl\ relationship updated.  The RMS
scatter in the \lhal-\lctm\ relation is $\sim$ 0.2 dex
\citep{GH05b}.  If we take the uncertainties on the \hal\ luminosity
as discussed in \citet{GH05b}, which are typically 0.13 dex, the
typical formal uncertainties on the BH masses are 0.14 dex based on
propagation of the measurement errors on the \hal\ luminosity and
width.  This random error does not include the important systematic
uncertainty in the normalization factor $f$. This is comparable to
the observable error for BH masses in Seyfert galaxies ($0.12-0.16$
dex) estimated by \citet{denney09} based on single-epoch data,
considering the effects of AGN variability and random measurement
errors.

In the discussion above, we followed GH07 and updated their recipe
to calculate \mbh.  The method is based on the \hal\ emission lines
only (not requiring a continuum measurement) and it allows us to
compare our results in a consistent way with previous work on SDSS
AGNs presented by GH07.  One particular point to consider is that
our mass estimates assume a normalization factor $f=0.75$ for the
virial masses, for consistency with GH07 and other previous work,
but this is not a unique choice for $f$. The virial factor has been
the subject of much discussion in the literature
\citep[e.g.][]{onken04, collin06}, and different $f$ values have
been used in various recipes to obtain \mbh\ estimates.
\citet{mcgill08} compared \mbh\ estimators based on different
emission-line and continuum measurements and showed that systematic
errors could be as large as $0.38 \pm 0.05$ dex.

We would like to examine how different recipes change the \mbh\
values of our sample and the \msigma\ relation that will be
discussed in Section~\ref{sec:msigma}.  Most of the recipes make use
of continuum luminosity, but we lack direct measurements of the
optical AGN continuum luminosity for our sample.  An alternative way
is to substitute \lctm\ with the broad \hal\ luminosity using the
empirical relationship given by \citet{GH05b}.  Some recipes rely on
measurements of \fwhm(\hbe), and to test those relationships we
substitute \fwhm(\hbe) with \fwhm(\hal) using the empirical
relationship between \hal\ and \hbe\ widths from \citet{GH05b}.  If
we use the recipe of \citet{vp06}, the masses would be higher by
about 0.25 dex.  The formalism presented by \citet{wjg09} gives
relatively larger BH mass than our estimator in the low mass end,
typically about 0.6 dex higher at BH mass of $\langle\log
M_{\mathrm{GH07}}/\msun \rangle = 6$.  They calibrated their BH mass
estimator by fitting the reverberation-based masses for 35 AGNs
using the continuum luminosity and \hbe\ \fwhm, as $\log \left( \mbh
/ \msun \right) = \alpha + \gamma \log \left( \lctm / 10^{42}
~\textrm{\ergs} \right) + \beta \log \left( \fwhm(\mathrm{H}\beta) /
10^3 ~\textrm{\kms} \right)$ and fixed $\gamma$ to be 0.5.  Their
fits yield values of $\alpha=1.39$ and $\beta=1.09$.  This value of
$\beta$ is less than $\beta=2$, which is commonly adopted for mass
estimators.  The 35 RM AGNs used in their fits mostly have BH masses
in the range from $10^7$ \msun\ to $10^9$ \msun, and $\alpha$ tends
to compensate for the mass difference in the fitting. This results
in larger BH masses in the low mass range, i.e., for objects below
about $10^7$ \msun.

There is some lower limit to the BH masses that we are able to
detect in SDSS data, due to a combination of factors.  A possibly
low BH occupation fraction in very low-mass galaxies would result in
a low detection rate.  The low luminosity of AGNs with small black
holes makes the AGNs hard to detect.  The host galaxy has to be
bright enough to be spectroscopically targeted by SDSS, and the S/N
of SDSS spectra is limited.  Moreover, the large aperture of SDSS
spectra can mix AGN emission with \ion{H}{2} regions and dilute the
AGN signal. Since we have identified the ``definite'' broad-lined
AGNs that have genuine broad \hal\ in our sample, we obtain a
cleaner sample that illustrates how low we can really go in
selecting objects with low \mbh\ using SDSS.
Our definite broad-lined sample with successful measurements of
\sigmastar\ has a median BH mass $\langle\mbh\rangle = 9.5 \times
10^5 \msun$, and a minimum of $2 \times 10^5 \msun$.

\subsection{\msigma\ Relation}
\label{sec:msigma}

\begin{figure}[t!]
\begin{center}
\scalebox{0.7}{\includegraphics{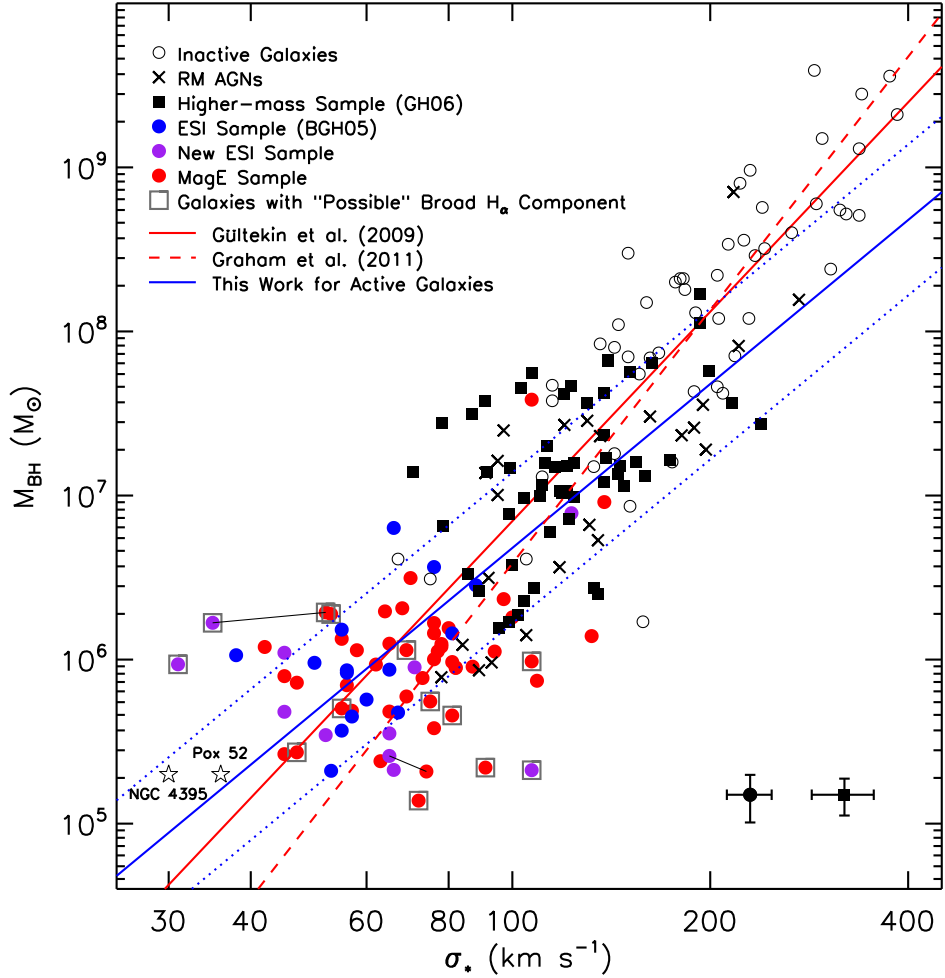}}
\end{center}
\caption{ The \msigma\ relation for massive black holes.  Filled
  dots represent new measurements from this work.  Red dots are the
  \mage\ sample, purple dots are the newly observed \esi\ objects, and
  blue dots are the objects originally from the BGH05 sample.  For
  objects observed with both \esi\ and \mage, both measurements are
  shown, connected by a short black line.  The filled dots
  over-plotted with grey squares represent the galaxies with
  ``possible'' broad \hal\ components.  Filled squares represent the 56
  higher-mass active galaxies selected from SDSS by \citet{GH06b},
  with BH masses updated with Equation~\ref{eq:mbh}.  Crosses
  represent literature data of 24 RM AGNs with stellar velocity
  dispersion measurements presented by \citet{woo2010}.  NGC 4395 and
  POX 52 \citep{filip03, peterson05, thornt08} are illustrated with
  stars. All the \mbh\ are scaled to the virial factor of $f=0.75$ or
  $f_{\sigma}=3$, in the virial mass $\mbh\ = f\rblr\Delta V^2 /G$
  (see text for details). Open circles represent objects with
  dynamically determined BH masses compiled by \citet{gult09}, and the
  red line is the \msigma\ relation derived by \citet{gult09}.
  Red dashed-line is the \msigma\ relation reported by \citet{grah11}.
  The blue solid line shows our best-fit \msigma\ relation for active
  galaxies with $\alpha=7.68\pm0.08$ and $\beta=3.32\pm0.22$, and
  the blue dotted lines show the intrinsic scatter of 0.46 dex.
  Typical errors in measurements for the sample
  in this work (filled dot) and GH06 sample (filled square) are
  shown on the bottom right (note that these do not include systematic
  uncertainties in the assumed $f$ factor). } \label{fig:bhsig}
\end{figure}

Figure~\ref{fig:bhsig} shows the \msigma\ relation for our sample of
AGNs with low BH masses, and for active galaxies with higher black
hole masses as well as nearby galaxies with direct
dynamical measurements.  The comparison sample of 56 active galaxies
with higher BH masses based on single-epoch spectroscopy were
selected from SDSS DR3 with $z \leq 0.05$ by \citet[][hereafter GH06
sample]{GH06b}, but the \mbh\ values have been re-calculated with
our updated mass recipe (Equation~\ref{eq:mbh}).  Literature data on
24 RM AGNs with stellar velocity dispersion measurements presented
by \citet[][and the references therein]{woo2010} were included.  The
\mbh\ for these RM AGNs were calculated from the virial products
(VPs) listed in Table 2 of \citet{woo2010}.  Since $\sigma_{\rm line}$
was used in VP, an isotropic velocity distribution gives
$f_{\sigma}=3$, assuming $\sigma_{line}=\fwhm/2$ \citep{onken04};
The ratio between \fwhm\ and $\sigma_{\rm line}$ is different for
different line profiles, i.e. $\fwhm/\sigma_{line}=2.35$ for a
Gaussian profile, and it varies around an average of 2
\citep{collin06}. Instead of the virial factor $f$ obtained by Woo
\etal, we assume $f_{\sigma}=3$ for consistency with the GH07
sample. This would decrease the masses by 0.24 dex compared to those
listed in their work.  We also included the well-known
intermediate-mass BHs in NGC 4395 with reverberation mass $\mbh =
(3.6 \pm 1.1) \times 10^5 \msun$ \citep{peterson05} and the velocity
dispersion with an upper limit $\sigmastar \leq 30$ \kms\ from
\citet{filip03}, and POX 52 with virial mass $\mbh \approx (3.1-4.2)
\times 10^5 \msun$ based on broad \hbe\ line width and \lctm\
\citep{barth04, thornt08} and velocity dispersion $\sigmastar = 36\pm5$ \kms\ from \citet{barth04}.
Masses for both objects were calculated utilizing the virial
coefficient $\langle f_{\sigma}\rangle=5.5$ \citep{onken04,
peterson04}, we also scaled down both masses by 0.26 dex for
consistency with other objects shown on the plot.  Note that the
virial normalization factor we assumed is arbitrary, and the derived
\mbh\ was decreased by about 0.26 dex compared to masses derived
using the $f$ factor from \citet{onken04}.  But the slope of
\msigma\ relation is independent of $f$.  Our low-mass sample
appears to smoothly follow the extension of the \msigma\ relation
for inactive galaxies, and there seems to be no strong change in
slope across the whole mass range including active and inactive
galaxies. We will quantify the slope for all the active galaxies
described above.

Assuming a log-linear form $\log \mbh = \alpha + \beta \log
(\sigmastar/200 ~\textrm{\kms})$, we fit the slope and zero point of
the \msigma\ relation for all the active galaxies with two
regression methods: the symmetric least-squares fitting method,
\emph{fitexy} \citep{press92} modified following \citet{tre02}, and
the maximum-likelihood estimate (MLE) method \emph{linmix\_err}
\citep{kelly07}, both implemented in IDL.  The former method
accounts for uncertainties in both coordinates as well as the
intrinsic scatter by adding a constant to the error in the dependent
variable, and solves for the best linear fit by minimizing the
reduced $\chi^2$ to unity.  \emph{Linmix\_err} uses a Bayesian
method to account for measurement errors and intrinsic scatter, and
computes a posterior probability distribution function of
parameters. Since there is no significant difference between the
results from the two regression methods, we will only quote results
from the MLE method \emph{linmix\_err}.  We obtain $\alpha = 7.68
\pm 0.08$ and $\beta = 3.32 \pm 0.22$, with $\epsilon_0 = 0.46 \pm
0.03$ dex. The slope is a bit flatter than that for nearby
galaxies, which have $\beta = 4.24 \pm 0.41$ \citep{gult09}, but
consistent with \citet{GH06b} and \citet{woo2010} who both showed
evidence of a shallower slope for active galaxies than the inactive
\msigma\ relation.  If we fix the slope $\beta$ to the best-fit
value of 4.24 for nearby galaxies, we obtain, for the full sample
of active galaxies, a zeropoint of $\alpha = 7.99\pm0.04$, which is
$-0.13\pm0.09$ offset from the value of $\alpha = 8.12\pm0.08$ from
\citet{gult09}.

\begin{figure}[t!]
\begin{center}
\scalebox{0.8}{\includegraphics{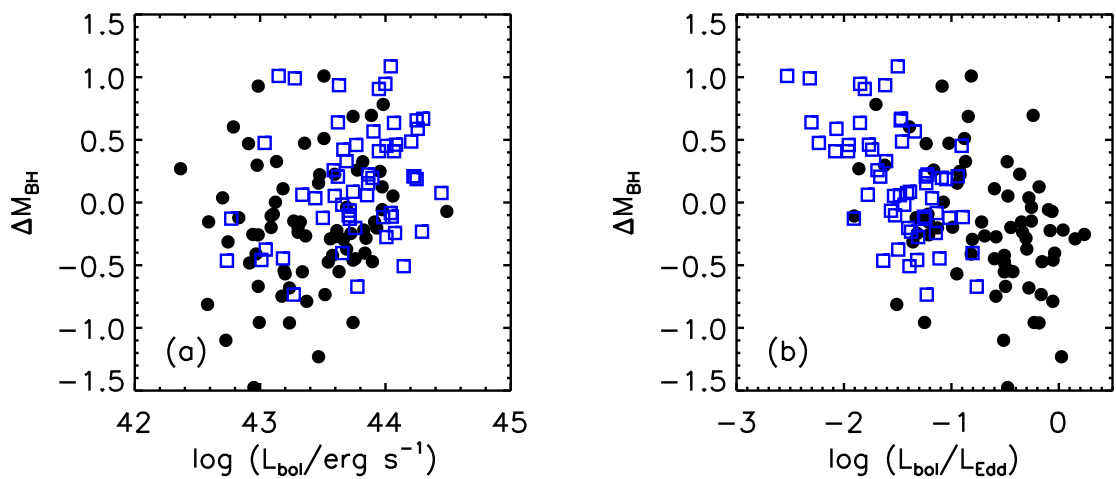}}
\end{center}
\caption{ The \msigma\ residual, $\Delta \mbh \equiv \log
  (\mbh/\msun) - \log(\mbh/\msun)_{\rm fit}$ vs. (a) bolometric
  luminosity, (b) \lledd.  Filled dots are the low-mass
  sample in this work. Blue squares are those 56 active
  galaxies selected from SDSS by \citet{GH06b}, with BH masses updated
  with Equation~\ref{eq:mbh}. } \label{fig:dmbh-lbol}
\end{figure}

We now consider residuals in the \msigma\ relation, defined as
$\Delta \mbh \equiv \log (\mbh/\msun) - \log(\mbh/\msun)_{\rm fit}$,
where $\log(\mbh/\msun)_{\rm fit}$ is calculated from \sigmastar, as
a function of the bolometric luminosity and Eddington ratio
(Figure~\ref{fig:dmbh-lbol}).  We follow GH07's method for
bolometric correction to estimate the bolometric luminosity from
\lhal, according to $\lbol = 2.34 \times 10^{44}(\lhal /
10^{42})^{0.86} ~\textrm{\ergs}$.  Among the 93 objects in our
sample, the median value of the Eddington ratio is
$\langle\lledd\rangle = 0.3$, where $\ledd \equiv 1.26 \times
10^{38} (\mbh/\msun) $.  The sample is dominated by objects
radiating at substantial fractions of their Eddington limits, since
the SDSS selection favors identification of the most luminous AGNs
in any mass range.  We find that the residual $\Delta \mbh$ is
significantly correlated with \lbol\ (Figure~\ref{fig:dmbh-lbol}a).
The Spearman rank correlation coefficient is $r_s = 0.33$, with a
probability $P < 10^{-3}$ that no correlation is present.  Moreover,
$\Delta \mbh$ shows a strong anti-correlation with \lledd\
(Figure~\ref{fig:dmbh-lbol}b, $r_s = -0.53$, $P < 10^{-9}$).  We are
wary to overinterpret these correlations, since the \lledd\ is
formally anti-correlated with \mbh\, and \mbh\ is correlated with
\lbol\ because both of them are deduced based on \lhal.

Note that for our \mbh\ estimates we are using propagated
measurement uncertainties based on the errors in \hal\ width and
luminosity, while the true uncertainties in \mbh\ are probably
dominated by the uncertainty in the BLR geometry and the chosen
value of $f$.  If we adopt 3$\sigma$ uncertainties in \mbh\ as
measurement errors and do the regression again, there is little
change in either the slope or zero point of the derived \msigma\
relation, but the intrinsic scatters decrease by about 35\% to
$\epsilon_0 = 0.28 \pm 0.05$ dex.  The regression parameters for
different samples and different uncertainties considered are listed
in Table~\ref{tab:reg}.  \citet{woo2010} recently reported the
\msigma\ relation for 24 RM active galaxies with BH mass $10^6 <
\mbh / \msun < 10^9$.  They obtained a slope $\beta\ = 3.55 \pm
0.60$ and intrinsic scatter $\epsilon_0 = 0.43 \pm 0.08$, which are
also consistent with our result.  In the following, we will examine
whether the intrinsic scatter varies as a function of mass.

We divide the active galaxies mentioned above into two data sets,
with lower and higher \mbh, and analyze the intrinsic scatter
$\epsilon_0$ for the two data sets.  The lower-\mbh\ data set
includes our sample and the Lick AGN Monitoring Project (LAMP)
sample presented by \citet{woo2010}.  The higher-\mbh\ data set
includes GH06 SDSS sample and 17 previous RM active galaxies
(collected also in \citet{woo2010}).  Assuming that the \mbh\ recipe
we used is equally valid at all masses (which is not necessarily the
case), we fit the two data sets and quantify the intrinsic scatter
to be $\epsilon_0 = 0.38 \pm 0.04$ dex for the low-\mbh\ sample
presented here and $\epsilon_0 = 0.40 \pm 0.04$ dex for the
higher-\mbh\ objects.  The scatter for low-mass data set slightly
changes when including NGC 4395 and POX 52, or excluding the LAMP
sample.  The scatter decreases a little to $\epsilon_0 = 0.34 \pm
0.04$ dex if we exclude the ``possible'' broad \hal\ objects.  The
derived scatters for the two data sets are consistent, indicating
that the intrinsic scatter in virial BH masses is not strongly
mass-dependent. We also examine the standard deviation of residuals
from the best-fit \msigma\ relation for the two data sets, and get
0.50 dex and 0.44 dex for the low and high mass range, respectively.

The intrinsic scatter in \msigma\ across the entire mass range is
about 0.46 dex, corresponding to a factor of $\sim$ 3.  That is
close to the intrinsic scatter for inactive galaxies, $\epsilon_0 =
0.44 \pm 0.06$ \citep{gult09}.  We will show in
\S\ref{sec:virf} that the intrinsic scatter persists or even
increases if we scale the active galaxies to follow the \msigma\
relation of nearby galaxies with direct measurements of BH masses
and then fit the relation for the
combined sample of the active galaxies and nearby galaxies.  Some of the
dispersion in the \msigma\ relation may result from different galaxy
morphological types following different \msigma\ loci, as recent
evidence has suggested \citep{hu08, ghb08, grali09, gult09,
gadotti09, greene10}. For example, the intrinsic scatter in \msigma\
relation of early-type galaxies (referring to elliptical galaxies
and S0 galaxies) is smaller than for late-type galaxies
\citep{grah08, gult09}. There is tentative evidence for a similar
trend in the \citet{GH04} sample, whose host galaxy morphologies
have been investigated by \citet{ghb08} using \hst\ images.  The
difference could either be due to unaccounted systematic error in
the \mbh\ measurements for spirals, or because the scatter in \mbh\
for spirals is actually larger, as might happen if the behavior of
BHs hosted by pseudobulges \citep[see][for a
review]{korken04} and classical bulges are different.  This has
been suggested by \citet{hu08}, who found that pseudobulges tend to
host lower \mbh\ than classical bulges at a given \sigmastar.
Similarly, a recent study by \citet{greene10} presents stellar
velocity dispersions for a sample of nine megamaser disk galaxies
with accurately measured BH masses in the range of $\sim10^7$
\msun. They find that the maser galaxies fall below the \msigma\
relation of elliptical galaxies defined by \citet{gult09}. Based on
morphology and stellar population properties, they speculate that
most of the nine megamaser galaxies are likely to contain
pseudobulges, providing further support for the idea that the
pseudobulge \msigma\ relation is offset below the relation for
classical bulges \citep{hu08, gadotti09}.  In \S\ref{morphology}
below, we investigate the possible dependence of the \msigma\
relation fit on host galaxy morphological type for our sample.

Simulations of massive BH growth indicates that the scatter in the
\msigma\ relation should increase toward the low-mass end.  In
models with high-mass seeds, galaxies evolve from well above the
present-day \msigma\ relation toward the relation, while galaxies
starting with low-mass seeds first lie far below the relation and
their BHs grow to move toward the relation \citep[][and references
therein]{vol2010}. Furthermore, \citet{vol2010} suggested that for
heavy seeds of $\sim 10^5$ \msun, the low-mass end of the \msigma\
relation will flatten to an asymptotic value.  In our sample, we do
not find clear evidence for increasing scatter in the \msigma\ relation
at low masses, and we can not discriminate among possible seed
models based on our sample. Increasing scatter at the low-mass end
should nevertheless be a generic result regardless of the seed
masses or the details of the BH growth mechanism.  As discussed by
\citet{peng07}, major mergers decrease the scatter of the
\mbh-\mgal\ relation because of a central-limit tendency of the
BH-to-galaxy mass relation to eventually converge to some mean
value.

Given the previous expectation that low-mass galaxies should show a
larger scatter in the \msigma\ relation than high-mass galaxies, as
well as the recent observational suggestions of an offset \msigma\
relation for pseudobulges relative to classical bulges, it remains
somewhat puzzling that the Type 1 AGN samples do not show an obvious
trend of either increasing scatter or changes in slope toward lower
masses. This appears to be the case both for the small sample of
AGNs with reverberation mapping \citep{woo2010}, and for the larger
sample of objects with single-epoch masses presented here.  It could
be that the random errors on individual AGN virial BH mass estimates
are so large that the scatter in the mass estimates (resulting from
variations in BLR kinematics, inclination, or other properties)
simply dominates over the intrinsic scatter in the \msigma\ relation
for these objects, preventing us from detecting mass-dependent
variations in the intrinsic scatter.  Direct stellar-dynamical
measurements of BH masses in nearby RM AGNs can be an important
cross-check on the derived masses.  However, while observations so
far do not reveal evidence for any dramatic offsets between
stellar-dynamical and reverberation masses, only a few objects are
presently amenable to both types of measurements \citep{davies06,
onken07, hicks08}.  For the vast majority of reverberation-mapped AGNs,
the angular size of the BH's gravitational sphere of influence is
too small to be resolved by either \hst\ or by ground-based telescopes
employing adaptive optics, and only a small number of broad-lined AGNs
are currently suitable targets for stellar-dynamical measurements of \mbh.

\subsection{Morphology: Barred versus Unbarred Disk Galaxies}
\label{morphology}

Recent work by \cite{grah11} has suggested that the \msigma\
relation is different for AGNs with different host galaxy
morphological types (i.e. barred disks or unbarred disks).  For our
sample, we can carry out a preliminary check for any offsets in the
\msigma\ plane as a function of host morphology.  The SDSS images
are not sufficient to carry out a full census of host galaxy types
for this sample, but there is morphological information for a subset
of our sample from an \hst\ Wide Field Planetary
Camera2 (WFPC2) imaging snapshot survey with the F814W filter
\citep{jiang11}.  Among the galaxies in our sample
having \hst\ images, there are 38 objects which are morphologically
classified as disk galaxies and which also have \mbh\ and velocity
dispersion measurements, including 12 barred and 26 unbarred
galaxies.  Only one galaxy from the GH06 higher-mass SDSS sample is
included in the \hst\ imaging survey; it is classified as an
unbarred disk galaxy. Including the reverberation-mapped AGN sample
compiled by \cite{woo2010}, as well as NGC 4395, and LAMP objects,
we have morphological information for a total of 63 disk galaxy AGN
hosts (25 barred and 38 unbarred). The morphological classifications
for the reverberation-mapped objects are from NED where available,
and otherwise adopted from classification by \cite{bentz09}, while
the classification for LAMP objects are provided by \citet[][in
preparation]{bentz11}.  If we remove the galaxies with a ``possible''
broad \hal\ component, there are 56 disk galaxies with morphological
information (22 barred and 34 unbarred). One caveat in
our division is that it is based on optical images, while it has
been suggested that some galaxies appear to be unbarred in optical
bands, but are very clearly barred when viewed in the infrared
\citep[e.g.][]{block91, mulreg97}.  Despite this caveat, we are
limited by the data presently available and we carry out a
preliminary examination based on the optical host-galaxy morphologies.

We re-do the regression fits for the \msigma\ relation as in
\S\ref{sec:msigma} for different subsamples, and the results are
listed in Table~\ref{tab:bar}.  The best fit of the \msigma\
relation for the full subsample of disk galaxies
($\alpha=7.50\pm0.13$, $\beta=3.04\pm0.30$) is consistent with that
for the entire sample of active galaxies described previously
($\alpha=7.68\pm0.08$, $\beta=3.32\pm0.22$).  Within the
uncertainties, the slope is consistent with the slope of
$4.05\pm0.83$ for 24 nonellipticals reported by \cite{gult09}.
Fixing the slope to 4.05 yields an intercept of
$\alpha=7.89\pm0.06$, which also agrees well with
$\alpha=8.01\pm0.16$ for nonellipticals \citep{gult09}.  The fit for
barred disks only gives $\alpha=7.81\pm0.27$ and
$\beta=4.13\pm0.72$, similar to the \msigma\ relation
($\alpha=7.80\pm0.10$, $\beta=4.34\pm0.56$) for barred inactive
galaxies presented by \cite{grah11}.

\begin{figure}[t!]
\begin{center}
\scalebox{0.7}{\includegraphics{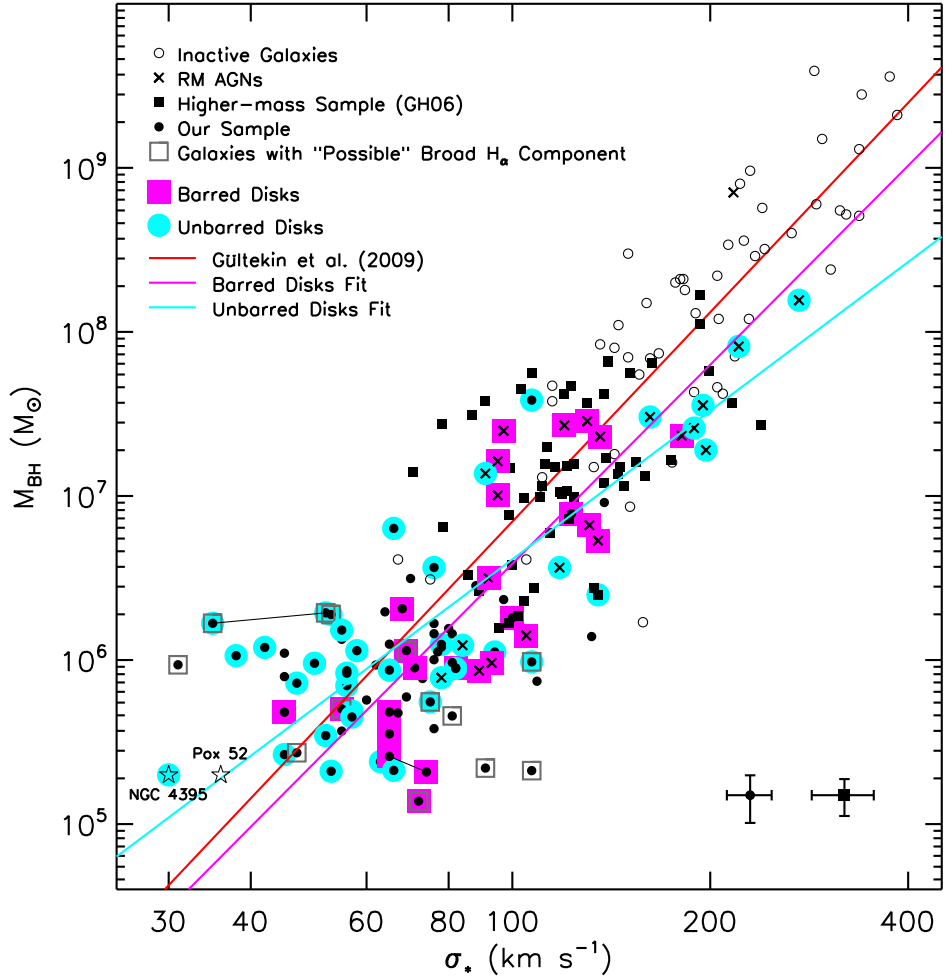}}
\end{center}
\caption{ The same as Figure~\ref{fig:bhsig}, with the morphological
classification of barred and unbarred disk galaxies shown.  Small
filled circles represent our sample with low-mass \mbh.  Large
magenta filled squares represent barred disk galaxies, while the
cyan circles denote unbarred disks. Magenta and cyan lines show the
best-fitting \msigma\ relation for the barred and unbarred
subsamples that excludes the $p$ objects with uncertain detections
of broad \hal, respectively. } \label{fig:bar}
\end{figure}

Figure~\ref{fig:bar} shows the morphological type for the disk
galaxies on the \msigma\ plot, and also the best fits of the
\msigma\ relations for barred and unbarred disks (see
Table~\ref{tab:bar}).  We consider our most reliable results to come
from the subset of our sample that excludes the $p$ objects with
uncertain detections of broad \hal.  For this subsample, we find
$\alpha=7.80\pm0.29$ for barred disks versus $7.52\pm0.15$ for
unbarred disks, and the slope $\beta=4.03\pm0.80$ for barred disks
versus $3.01\pm0.33$ for the unbarred subset.  Thus, the barred
disks have a marginally larger zeropoint and steeper slope than
unbarred disks, but the difference is not significant given the
substantial uncertainties on the fits.  We also fit the \msigma\
relation with the slope fixed to that of inactive galaxies, $4.24$
\citep{gult09}, as an additional test for any offset.  With fixed
slope, the distinction between the barred and unbarred subsamples
almost vanishes (see Table~\ref{tab:bar}). Alternatively, if we were
to adopt the high slope value proposed by \cite{grah11}
($\beta=5.13$), then the zeropoints would be all increased by about 0.35.

\subsection{Disk Inclination}
\label{sec:incl}

We have measured \sigmastar\ for our sample from the \esi\ and \mage\
spectra obtained with slit widths of 0\farcs75 and 1\arcsec,
respectively.  At the median redshift ($\langle z \rangle=0.08$, or
luminosity distance $D_L$=300 Mpc) for our sample, 1\arcsec\
corresponds to a scale of 1.5 kpc.  If the bulge is not the dominant
component of the host galaxy over that scale, then the slit spectra
will be contaminated by the light from the disk, and the measured
\sigmastar\ will include contributions from the rotational velocity of
stars orbiting on the disk.  Specifically, \sigmastar\ will be
increased artificially in an edge-on disk, and might be slightly
decreased in a face-on disk.  In order to examine the effect of disk
inclination on our sample, we use the axis ratio $b/a$ to indicate
inclination, where $b/a$ close to 1 implies a face-on
(low-inclination) system and $b/a$ close to 0 denotes an edge-on
(high-inclination) system.  The disk axis ratio is obtained from fits
to \hst\ images.  For the subset of our sample having \hst\ imaging,
the axis ratio measurements are based on the analysis of the WFPC2 snapshot
imaging survey data \citep{jiang11}.
For reverberation-mapped AGNs, axis ratios measured from \hst\
imaging data are taken from \citet[][their Table 4]{bentz09} and
from \citet[][in preparation]{bentz11}.  We
divide the sample into three axis-ratio bins: $b/a > 0.88$,
$0.72<b/a<0.88$, and $b/a < 0.72$.

\begin{figure}[h]
\begin{center}
\scalebox{0.6}{\includegraphics{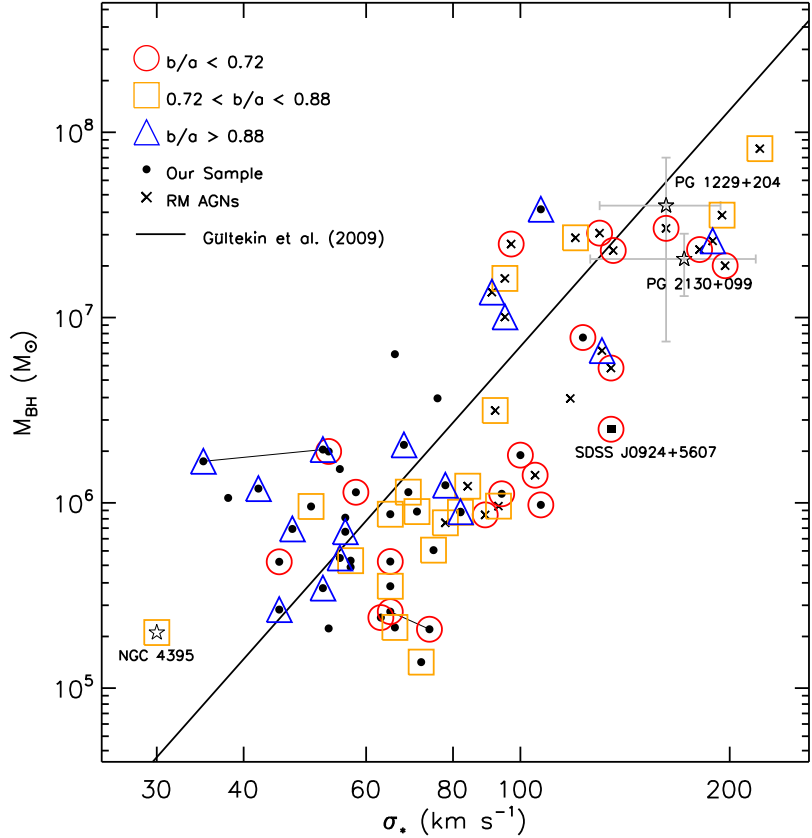}}
\end{center}
\caption{The same as Figure~\ref{fig:bhsig}, enlarged to highlight
the active galaxies classified as disk galaxies.  Small filled
circles, squares and crosses represent the same as
Figure~\ref{fig:bhsig}.  Three subsamples of high-, medium- and
low-inclination objects are overplotted with red open circles,
orange open squares and blue triangles, respectively.  The \msigma\
relation from G\"{u}ltekin \etal\ is shown for comparison.  Only one
galaxy (SDSS J0924+5607) from the GH06 higher-mass SDSS sample is
observed in the \hst\ imaging survey.  Two extra PG objects are also
shown as stars.  For PG 1229+204 and PG 2130+099, the BH masses are
$4.0\times 10^7$\msun\ and $2.1\times 10^7$\msun\ \citep[virial
product from][]{peterson04, grier08} under the assumption of
$f_{\sigma}=3$ as discussed in \S\ref{sec:msigma}, and velocity
dispersion are $162\pm32$ \kms\ and $172\pm46$ \kms\ \citep{dasyra07},
respectively.  Their axis ratios of disk components are 0.62
and 0.55, respectively, and they are classified as high-inclination
objects.} \label{fig:inc}
\end{figure}

\begin{figure}[h]
\begin{center}
\scalebox{0.5}{\includegraphics{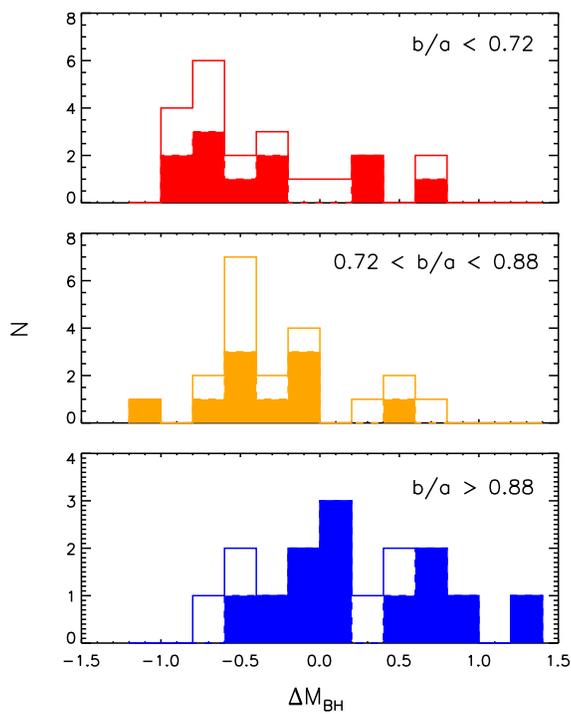}}
\end{center}
\caption{Distribution of $\Delta$\mbh\, defined as the \mbh\
deviation from the best-fit \msigma\ relation from G\"{u}ltekin
\etal, in three bins of disk inclination with the axis ratio range
shown at the top right in each panel.  From top to bottom is high-
(close to edge-on), medium- and low-inclination (close to face-on).
In each panel, the open and filled histograms show the distributions for
all AGNs having host galaxy axis ratios (including the
reverberation-mapped objects) and the subset from only our SDSS
sample, respectively. } \label{fig:dist_dmbh}
\end{figure}

Figure~\ref{fig:inc} shows the disk galaxies compiled in
\S\ref{morphology} in the \msigma\ plot, labeled by axis ratio.  The
low-inclination objects (large blue triangles) lie apparently to the
left of the medium- (large orange squares) and high-inclination (large
red circles) objects.  We define $\Delta$\mbh\ as the \mbh\ deviation
from the best-fit \msigma\ relation reported by \cite{gult09}.  The
distributions of $\Delta$\mbh\ for the three bins are shown in
Figure~\ref{fig:dist_dmbh}.  A K-S test on the distributions of
$\Delta$\mbh\ for the low- and high-inclination subsamples yields a
probability of 0.011, indicating that the distribution of the offset
$\Delta$\mbh\ is likely to be significantly different for these two
subsamples.  Two extra PG objects (PG 1229+204 \&\ PG2130+099) with
\mbh\ and \sigmastar\ measurements are also shown on
Figure~\ref{fig:inc}.  Their axis ratios of disk components are 0.62
and 0.55, respectively, and they are classified as high-inclination
objects.  Including these two objects in the K-S test above does not
alter the result.  We also carry out fits to \msigma\ relation for
each subsample, and list the best-fit parameters in
Table~\ref{tab:inc}.  With the slope fixed to 4.24, the
low-inclination subsample shows offsets from the other two subsamples
by more than 0.4 dex.  This gives additional evidence that the
low-inclination and high-inclination systems are significantly offset
in the \msigma\ plane, such that the more highly inclined host
galaxies tend to have higher \sigmastar\ at a given value of \mbh.
The offset is consistent with the expectation that disk rotation
contributes more significantly to the measured values of \sigmastar\
in the more highly inclined systems.  This may be one source of error
contributing to the scatter in the relation, and it may also introduce
a bias to the zeropoint of the relation.  Although it would be
preferable to plot an \msigma\ relation that did not include this
possible bias, there is no simple way to correct for the rotational
contribution to the velocity dispersions from our data.

\subsection{Virial Normalization Factor}
\label{sec:virf}

In the discussions above, we have adopted an arbitrary virial factor
$f=0.75$ for consistency with most previous work on the
SDSS-selected GH07 sample.  Alternatively, we can also fit for the
value of $f$ that brings our sample into the best agreement with the
local \msigma\ relation, following the same method used by
\cite{onken04} and \cite{woo2010} for reverberation-mapped AGNs. The
values reported by \cite{gult09} are $\alpha=8.12\pm0.08$ and
$\beta=4.24\pm0.41$.  Fixing the slope to 4.24 for our full sample,
we obtain $\alpha=7.99\pm0.04$ for $f=0.75$ as presented in
\S\ref{sec:msigma}.  Therefore, the mean virial factor determined
from this fit would be $\langle f \rangle =1.01\pm0.21$.
Alternatively, fitting our sample to the \msigma\ relation from
\citet{grah11} would give a lower value of $\langle f \rangle
=0.52\pm0.08$.  If we take the
subsample of the 15 low-inclination galaxies above in \S\ref{sec:incl}
and scale to the \msigma\ relation from G\"{u}ltekin \etal, we get
$\langle f \rangle =0.56\pm0.17$.  These results are reasonably compatible with
previous calibrations of $\langle f \rangle$ for
reverberation-mapped AGNs, taking into account the fact that our
work uses the \fwhm\ as the measure of line width while the
reverberation-based mass scale is based on the second moment of the
\hbe\ line \citep[e.g.,][]{onken04, woo2010}. Fundamentally, the
calibration of $\langle f \rangle$ via the \msigma\ relation is
still limited by small-number statistics, and improving the
determination of the virial normalization factor will require
increasing the number of AGNs having high-quality reverberation
mapping data as well as the number of spiral galaxies having direct
measurements of \mbh\ from spatially resolved dynamics, and equally
important, obtaining accurate \sigmastar\ measurements for them.

For completeness of our analysis, we fit the \msigma\ relation
for the combined sample of active and nearby galaxies.  First we use
the virial factor derived above to recompute the BH masses for the active
galaxies.  Then we re-do the regression fit for the combined sample.
Using the virial factor $\langle f \rangle=1.01$ derived above, which is based on
normalizing the AGNs to the \msigma\ relation of G\"{u}ltekin \etal, we
combine the AGNs with the 49 nearby galaxies with direct dynamical
measurements of BH masses presented by the same authors.  Then
the regression for the combined sample yields $\alpha=8.02\pm0.05$ and
$\beta=3.81\pm0.16$ with an intrinsic scatter of $\epsilon_0=0.46\pm0.03$.
If the alternative $\langle f \rangle=0.52$ is used, then the fit yields
a zero point of $7.89\pm0.06$ and a slope of $4.14\pm0.16$ with an intrinsic scatter of
$0.49\pm0.03$.\footnote{
Instead we could choose $\langle f \rangle=0.52$ based on normalizing the AGNs to the
\msigma\ relation of \cite{grah11}, and then combine the AGNs with
the 64 galaxies with direct BH mass measurements listed by Graham et al.  Assigning a 10
percent error to the \sigmastar\ for the 64 galaxies as Graham \etal\ suggested,
the regression for the combined sample yields $\alpha=7.93\pm0.05$ and
$\beta=4.30\pm0.16$ with $\epsilon_0=0.48\pm0.03$.  Assigning a 5 percent error to
the \sigmastar\ instead will yield almost the same results. }
The intrinsic scatter for the combined
sample is the same as or even larger than that for active galaxies only.
We note that the slope and the intrinsic scatter depend on how we
select the virial factor, because the active galaxies are scaled to
the local \msigma\ relation first and then included for
the fit.

\section{Narrow Emission-line Properties}
\label{sec:nel}

\subsection{Stellar vs. Gas Velocity Dispersion}
\label{sec:stelvsgas}

It is well-known that the velocity dispersion of the ionized gas in
the narrow-line region (NLR), most often measured from the
[\ion{O}{3}] $\lambda5007$ \AA\ line, correlates with the stellar velocity
dispersion of the host galaxy bulge, suggesting that gravity dominates
the global kinematics of the NLR for active galaxies
\citep[e.g.][]{whit85, nels96, boroson03, GH05a, bian06, komossa07,
barth08}.  The high-resolution \esi\ and \mage\ data provides an
opportunity to examine this relationship in galaxies of small
\sigmastar.  In addition to the [\ion{O}{3}] line, we also consider
lines from lower ionization species such as [\ion{N}{2}] $\lambda6583$ \AA\
and the [\ion{S}{2}] $\lambda\lambda 6716, 6731$ \AA\AA\ doublet, and check
their relationship to \sigmastar\ for our sample.

For the [\ion{O}{3}] $\lambda\lambda4959,5007$ \AA\AA\ lines, each was
fitted by a double-Gaussian model, representing a core and often
(but not always) a blueshifted wing component.  Generally the core
component has a higher peak flux than the wing component.  In some
cases, the two-component model sometimes over-fits noise
fluctuations near the line peak, resulting in individual Gaussian
components with very narrow widths.  We set an empirical criterion
for the separation of the two components ($> 80$ \kms) to define the
blue- and red-shifts.  The blue- or red-shifted wing component was
removed from the $\fwhm_{core}$ measurements if the separation was
$> 80$ \kms.  For the other cases, where the centroids of the two
components were separated by less than 80 \kms, we measured the
\fwhm\ from the [\ion{O}{3}] profile and took the resulting value to
be $\fwhm_{core}$.  Detailed studies of some galaxies
where the NLR is spatially resolved reveal the possibility that the
entire [\ion{O}{3}] line may be outflowing \citep[][and references
therein]{fisch11}.  We use the relative velocity of the narrow component
of \hbe\ to help determine the core component, since the Balmer
recombination lines are less affected by the outflow when present.

\begin{figure}[hp]
\begin{center}
\scalebox{0.8} {\includegraphics{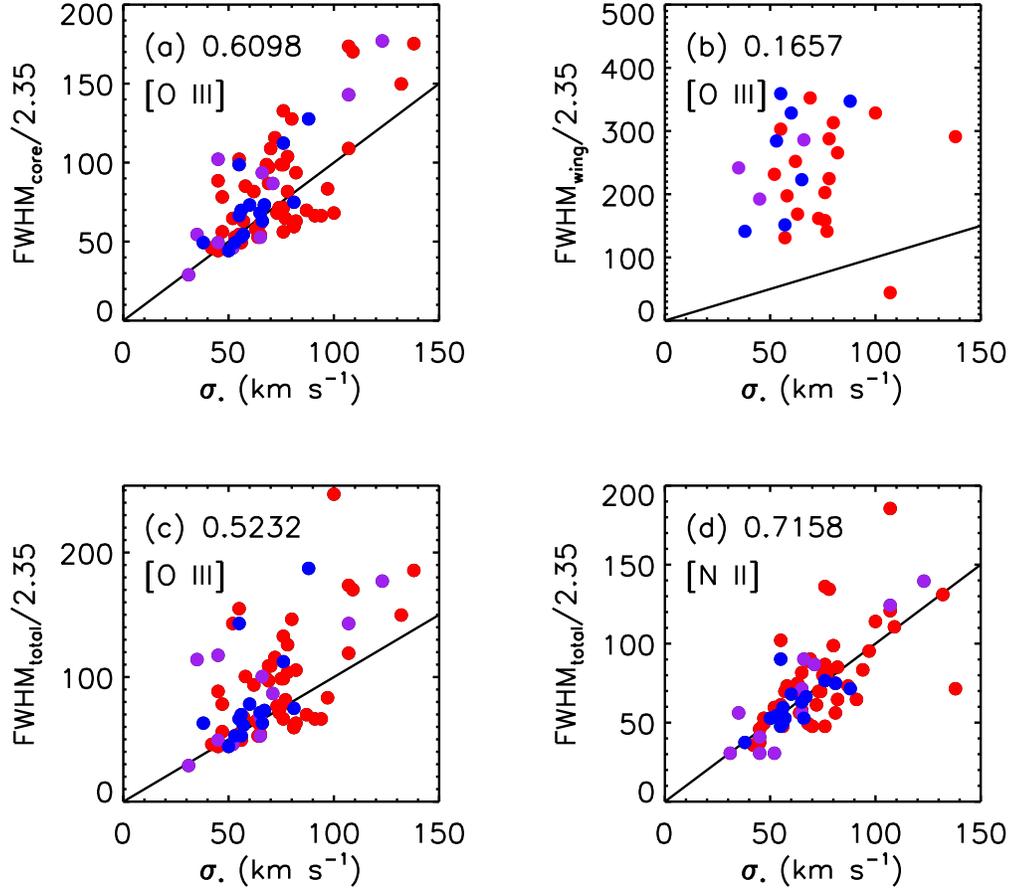}}
\end{center}
\caption{\fwhm/2.35 vs.\ \sigmastar\ for: (a) the core component of
[\ion{O}{3}] after removing the shifted wing component; (b) the
blue-shifted or red-shifted wing component of [\ion{O}{3}]; (c) the
overall profile of [\ion{O}{3}]; and (d) the overall profile of
[\ion{N}{2}].  The colors of plot symbols are as in
Figure~\ref{fig:bhsig}.  The number at the top of each panel gives
the Spearman rank correlation coefficient for the plotted data.  The
solid line in each panel represents $\fwhm/2.35 = \sigmastar$. }
\label{fig:gas-sig}
\end{figure}

\begin{figure}[hp]
\begin{center}
\scalebox{0.8}{\includegraphics{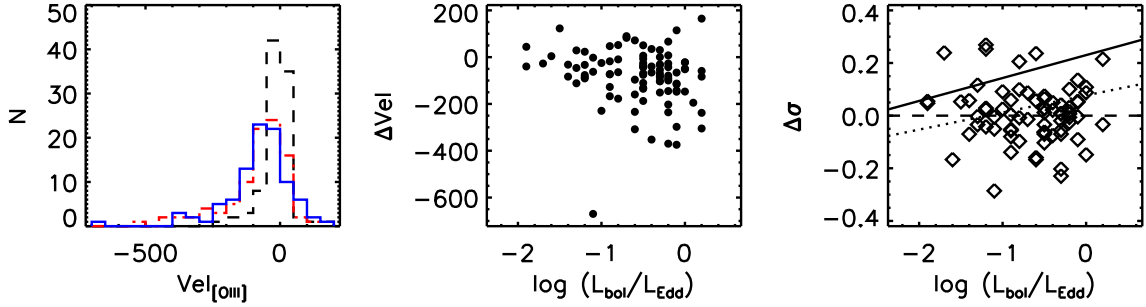}}
\end{center}
\caption{\textit{Left panel}: Distribution of velocity shifts
relative to the stellar absorption system for the core component
(black dashed) and the wing component (red dot-dashed) of [\ion{O}{3}]; Distribution
of $\Delta$Vel (\kms), defined to be the relative velocity-shift between
the wing to core components, is shown in blue solid histogram.
\textit{Middle panel}: $\Delta$Vel as a function
of Eddington ratio. The object with $|\Delta {\rm Vel}| > 600$ \kms\
is SDSS J145045.54$-$014752.8, showing obvious double-peaked
features in narrow lines. \textit{Right panel}: Distribution of the
residuals $\Delta\sigma\equiv \log \sigma_g - \log \sigmastar$
($\sigma_g \equiv $ \fwhm/2.35) versus Eddington ratio,
$L_{bol}/L_{Edd}$, for [\ion{N}{2}] $\lambda6583$ \AA.  The dashed
line denotes $\sigma_g = \sigmastar$.  The solid and dotted lines
show the relations found by \citet[][Equation 3]{ho09} and
\citet[][Equation 4]{GH05a}. } \label{fig:eddrt-sig}
\end{figure}

Figure \ref{fig:gas-sig} illustrates the relationships between
\fwhm([\ion{O}{3}])/2.35 and \sigmastar\ for the core and wing
components of the [\ion{O}{3}] line as well as for the entire
[\ion{O}{3}] profile (Figure~\ref{fig:gas-sig}c).  The \fwhm\ of the
[\ion{O}{3}] core correlates with \sigmastar, with a Spearman rank
correlation coefficient $r_s=0.61$, and a probability
$P_\mathrm{null}$ less than $10^{-4}$ for the null hypothesis of no
correlation (Figure~\ref{fig:gas-sig}a).  The relationship between
the \fwhm\ of the entire [\ion{O}{3}] profile and \sigmastar\ has a
larger scatter, consistent with previous results
\citep[e.g.][]{nels96, GH05a, barth08}.  This illustrates that if we
use the \fwhm\ of [\ion{O}{3}] as a proxy for \sigmastar\ in active
galaxies, the high-velocity wing component should be removed from
the profile. The velocity shifts between the centroids of the wing
to core components, defined as $\Delta$Vel, are mainly negative
(Figure~\ref{fig:eddrt-sig}, left panel), representing blue-shifts
of the wing components, while the centroid velocity of the core
components are generally close to the systemic velocity measured
from the stellar absorption lines.  If the blue-shifted wing
component of [\ion{O}{3}] is driven by winds or outflows
\citep{whit85, komossa08}, then the strength or velocity offset of
the wing component could correlate with some measure of AGN
activity, such as the Eddington luminosity ratio, \lledd.  Thus we
investigate how the equivalent width (EW) ratio of the [\ion{O}{3}]
wing to core components correlates with Eddington ratio, but we find
no correlation between them.  We find that $\Delta$Vel correlates
with \lledd\ only weakly ($r_s=-0.24$, Figure~\ref{fig:eddrt-sig},
middle panel).  We also find that $\Delta$Vel correlates well with
the \fwhm\ of the wing component ($r_s=0.56, P_{null}<0.001$) and
less strongly with the \fwhm\ of the core component ($r_s=0.41,
P_{null}=0.03$).  This confirms that the previously established
correlation between [\ion{O}{3}] blueshift and line width continues
to hold in this low-mass regime.

We also examine the correlation between gas and stellar velocity
dispersions for the [\ion{S}{2}] doublet and the [\ion{N}{2}]
$\lambda6583$ \AA\ emission line, which has been decomposed from broad
\hal.  Since we use the [\ion{S}{2}] profile as the model for the
narrow lines for most cases (discussed in Section~\ref{sec:emline}),
we can use either the [\ion{S}{2}] or [\ion{N}{2}] lines.  The
[\ion{N}{2}] line is adopted here because there are a
few cases for which [\ion{S}{2}] is too weak to measure its width
reliably.  The \fwhm\ is measured from the overall, combined profile
of gaussians fitted to each narrow emission line.  As shown in Figure
\ref{fig:gas-sig}, the \fwhm\ of [\ion{N}{2}] correlates well with
\sigmastar, with a Spearman rank correlation coefficient $r_s = 0.72$
and $P_{null}<10^{-4}$.  We follow \citet{ho09} to define the gaseous
velocity dispersion $\sigma_g \equiv $ \fwhm/2.35 and the residual of
$\sigma_g-\sigmastar$ to be $\Delta\sigma\equiv \log \sigma_g - \log
\sigmastar$.  While \citet{ho09} finds a correlation between
$\Delta\sigma$ and Eddington ratio for nearby active galaxies, we do
not see such a correlation in our sample (Figure~\ref{fig:eddrt-sig},
right panel).  Our sample mostly lies below the relation found by
\citet{ho09}, but is generally consistent with \citet{GH05a}. The
objects scatter around $\Delta\sigma=0$ ($\sigma_g = \sigmastar$),
regardless of increasing Eddington ratio.

\subsection{Emission-line Diagnostics}
\label{sec:bpt}

\begin{figure}[hp]
\begin{center}
\scalebox{0.8}{\includegraphics{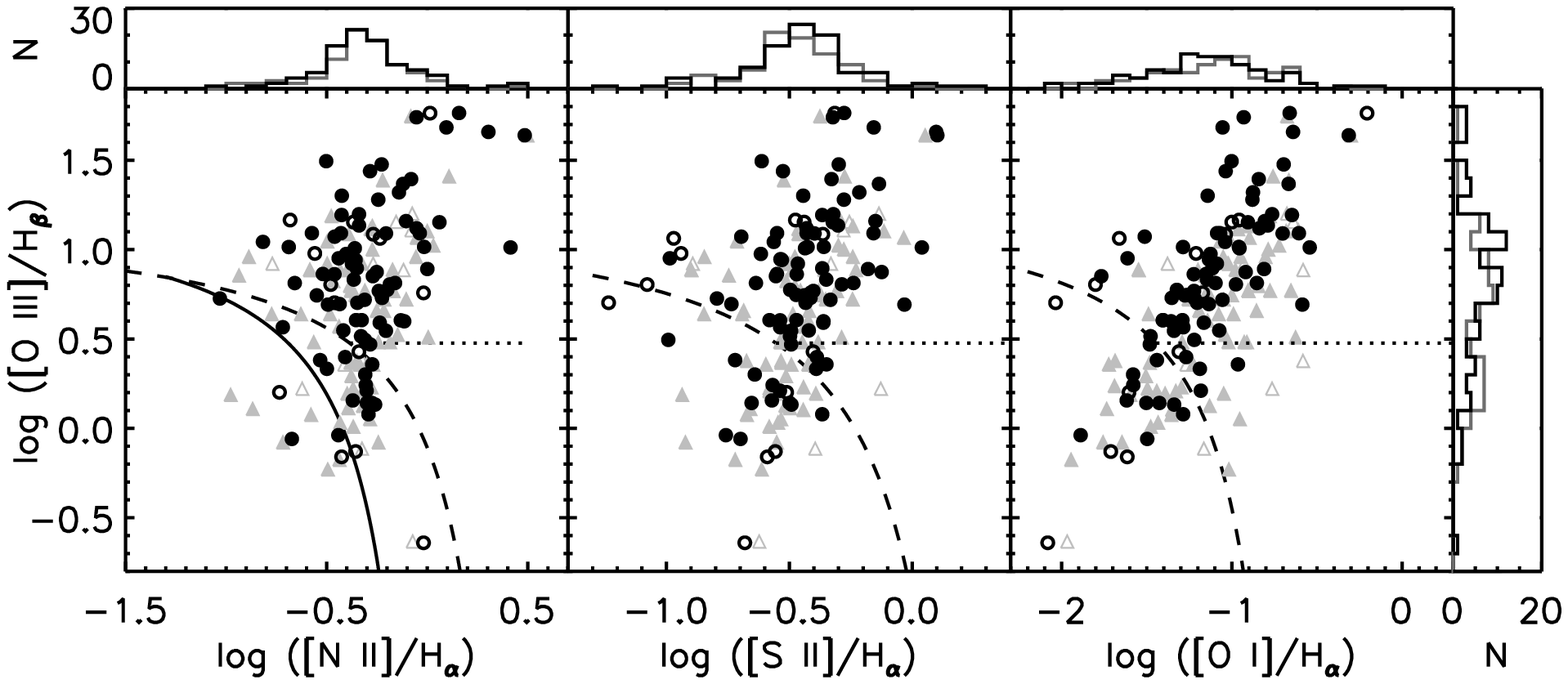}}
\end{center}
\caption{Diagnostic diagrams for [\ion{O}{3}]/\hbe\ vs.\
[\ion{N}{2}]/\hal, [\ion{S}{2}]/\hal\ and [\ion{O}{1}]/\hal.  Filled
circles represent objects with definite broad \hal, and open circles
represent objects classified as having possible broad \hal. Grey
triangles show the line ratios measured from the SDSS data by GH07
(open triangles represent their candidate sample). Dashed curves are
the maximum starburst lines from \citet{kewley06}; the solid curve
is the pure star formation boundary \citep{kauff03}; The horizontal
dotted line marks [\ion{O}{3}]/\hbe$ = 3$. } \label{fig:bpt}
\end{figure}

We have measured emission lines of \hbe, [\ion{O}{3}] $\lambda5007$
\AA, \hal, [\ion{N}{2}] $\lambda6583$ \AA, and [\ion{S}{2}]
$\lambda\lambda6716,6731$ \AA\AA, which enable us to plot our sample on
line-ratio diagnostic diagrams \citep[e.g. BPT diagrams;][]{bpt81,
vo87}, as shown in Figure~\ref{fig:bpt}, and to compare our
measurements with the SDSS measurements obtained with larger
spectroscopic apertures (fiber diameter of 3\arcsec).  The flux of
[\ion{O}{1}] $\lambda6300$ \AA\ was also measured from the
starlight-subtracted spectra by fitting with a single Gaussian.  From
the new data (with aperture sizes of 0\farcs75 $\times$ 1\farcs0 and
1\farcs $\times$ (1\farcs5 $\sim$ 3\arcsec) for the \esi\ and \mage\ data,
respectively), we see that a few objects move out of the \ion{H}{2}
region of the BPT diagrams and into the Seyfert region above the
maximum starburst line \citep{kewley06}.  With the small-aperture
data, most objects lie in the Seyfert region of the diagram, with some
objects extending to relatively low values of [\ion{N}{2}]/\hal\ which
are indicative of metallicity lower than typical for classical
Seyferts \citep{groves06}.

We investigate whether the line ratios measured from the
small-aperture spectra systematically change relative to
measurements based on the SDSS spectra.  The distributions of line
ratios are shown on the top and right panels in
Figure~\ref{sec:bpt}.  These distributions for the low-ionization
ratios [\ion{N}{2}]/\hal, [\ion{S}{2}]/\hal, and [\ion{O}{1}]/\hal\
only change slightly as a function of aperture size, with systematic
offsets less than 0.08 dex between the median values for the
small-aperture and SDSS measurements.  This could be the effect of
various parameters, due to the complex NLR conditions, including
metallicity, ionization parameter, electron density, and dust
reddening.  But the [\ion{O}{3}]/\hbe\ ratio measured from the
small-aperture data is systematically 0.3 dex higher than the values
measured from the SDSS fiber aperture.  The simplest interpretation
is that in the larger SDSS aperture the emission lines included a
larger contribution of emission from \ion{H}{2} regions, which would
have lower [\ion{O}{3}]/\hbe\ than the AGNs.  This would also imply
that the host galaxies are actively star-forming, and we may be
witnessing growth of the host galaxy bulge or pseudobulge coevally
with the BH growth.  Integral-field observations at high angular
resolution would be particularly useful for examining the
distribution and luminosity of \ion{H}{2} regions in the host
galaxies.

These line-ratio diagrams are generally used to discriminate between
star-forming galaxies and AGNs \citep[e.g.][]{ho97, kauff03, hao05,
barth08}.  However, as noted by GH07, the broad-line AGNs do not lie
exclusively above the maximum starburst line of Kewley \etal\ on the
BPT diagrams, and some objects even fall below the empirical line of
\citet{kauff03} in the region of pure star-forming galaxies.  That
is, judging by the narrow-line ratios alone, these objects would not
be classified as AGNs.  For these objects, the detection of the
broad \hal\ emission is the primary clue that an AGN is present, and
if the broad emission lines were obscured by a dusty parsec-scale
torus or by dust lanes in the host galaxy then these objects would
not have been identified as AGNs at all.

We explore whether the objects which fall below the maximum
starburst line are unusual in terms of any of the measured
parameters including \sigmastar, $\sigma_{gas}$ (indicated as
\fwhm([\ion{N}{2}])/2.35), the flux ratio between [\ion{O}{3}] wing
and core, the \ion{Fe}{2} strength (relative to [\ion{O}{3}]),
\fwhm(\hal), or the flux ratio between the narrow and broad
components of \hal\ ($F_{n/b}$).  We find that this sub-population
of objects does not exhibit any unusual properties in terms of these
parameters, except in $F_{n/b}$, which shows a higher relative
strength of the narrow components in this sub-population.  Further
investigation shows the median redshift of the sub-population
($z\sim0.1$) is larger than the whole sample ($z\sim0.08$).  Thus
the higher $F_{n/b}$ ratio could be explained by the dilution of
extended star formation contained in the aperture.  It is
unfortunately difficult to derive any detailed information on the
stellar populations of the hosts from the spectra because of the
strong AGN contribution to the spectra. We examined the \esi\ and
\mage\ spectra, finding that higher-order Balmer absorption of
significant strength is visually apparent in 18 objects (about 20\%
of the sample), of which about 10 objects are in the sub-population
below the maximum starburst line. The absorption features imply
contributions from A stars.  Thus, compared with the sample as a
whole, a larger fraction of objects in the \ion{H}{2} sub-population
appear to have significant contributions from intermediate-age
stellar populations.

\section{Conclusions}
\label{sec:con}

We have obtained new echelle spectroscopy of 76 Seyfert 1 galaxies
selected to have low-mass black holes based on their broad-line
widths and \lhal\ luminosity, and we obtain reliable measurements of
\sigmastar\ for 56 galaxies.  Including the previous 17 Seyfert
galaxies presented by \citet{barth05}, our low-mass AGN sample
consists of 93 galaxies, of which 71 have measured velocity
dispersions.  The velocity dispersion ranges from 31 to 138 \kms,
and the data tend to lie on the extrapolation of the \msigma\ relation of
inactive galaxies.  We find that the intrinsic scatter in virial \msigma\ relation
is not strongly mass-dependent.  Combining our results with 56 SDSS active
galaxies \citep{GH06a} and reverberation-mapped AGNs from the literature
with \sigmastar\ measurements and \mbh\ estimation, as well as the
two well-known intermediate-mass BHs in NGC 4395 and POX 52, we carry
out new fits to the \msigma\ relation, finding a zero point of
$\alpha\ = 7.68 \pm 0.08$ and slope $\beta\ = 3.32 \pm 0.22$ for the
\msigma\ relation in the form of $\log \mbh = \alpha + \beta \log
(\sigmastar/200 \textrm{\kms})$, for an assumed virial normalization
factor of $f=0.75$.  The AGN \msigma\ relation has an intrinsic
scatter of $0.46\pm0.03$ dex, comparable to the intrinsic scatter of the
\msigma\ relation observed for inactive galaxies.  Among our
low-mass BH sample with definite broad \hal\ components, we find
that the lower limit of BH masses detectable in this SDSS sample is
$2 \times 10^5 \msun$ (for $f=0.75$).  We do not find a significant
offset or slope difference in the \msigma\ relation between the
subsamples of barred and unbarred disk galaxies.  We do
find that the disk galaxies with high inclination angles (edge-on
systems) show a mild offset from the face-on systems in the \msigma\
relation.  The rotation of the disk in the edge-on systems may
artificially increase the measured \sigmastar, which may partly
introduce a scatter in the relation.  We also confirm that the
narrow emission lines [\ion{N}{2}] $\lambda6583$ \AA, [\ion{S}{2}]
$\lambda\lambda6716,6731$ \AA\AA, and the core of [\ion{O}{3}]
$\lambda5007$ \AA\ (with the blue-shifted wing removed) have velocity
dispersions which trace the stellar velocity dispersion well,
confirming that the forbidden emission-line widths can be used as a
useful proxy for the stellar velocity dispersion in low-mass AGNs.

\acknowledgements

We thank the anonymous referee for comments and suggestions that
helped to improve the manuscript.  We are grateful to Virginia
Trimble, Tinggui Wang, and Xiaobo Dong for advice and discussions
during the course of this work, George Becker for his reduction
package (\magerdc) for \mage\ echelle spectra, and Thomas Matheson
for helpful discussion on flux calibration.  T.X. acknowledges
support from the Chinese Scholarship Council during her visit at UC
Irvine, where main part of this work was carried out. Research by
A.J.B. was supported by NSF grant AST-0548198. Some of the data
presented herein were obtained at the W.M. Keck Observatory, which
is operated as a scientific partnership among the California
Institute of Technology, the University of California and the
National Aeronautics and Space Administration. The Observatory was
made possible by the generous financial support of the W.M. Keck
Foundation. The authors wish to recognize and acknowledge the very
significant cultural role and reverence that the summit of Mauna Kea
has always had within the indigenous Hawaiian community.  We are
most fortunate to have the opportunity to conduct observations from
this mountain.  Funding for the SDSS has been provided by the Alfred
P. Sloan Foundation, the Participating Institutions, the National
Science Foundation, the U.S. Department of Energy, the National
Aeronautics and Space Administration, the Japanese Monbukagakusho,
the Max Planck Society, and the Higher Education Funding Council for
England. The SDSS Web Site is http://www.sdss.org/.  This research
has made use of the NASA/IPAC Extragalactic Database (NED) which is
operated by the Jet Propulsion Laboratory, California Institute of
Technology, under contract with the National Aeronautics and Space
Administration.

\clearpage

\begin{deluxetable}{llcccccccccccc}
\rotate
\tablewidth{0pt}
\tablecaption{Observations and Measurements}
\setlength{\tabcolsep}{0.02in}
\tablecolumns{14}
\tabletypesize{\scriptsize}
\tablehead{
\colhead{SDSS Name} & \colhead{Flag} & \colhead{z} & \colhead{Exp.}
& \colhead{S/N} & \colhead{$\sigma$(\mgb)} & \colhead{$\sigma$(Fe)}
& \colhead{$\sigma$(\caii)} & \colhead{\sigmastar} &
\colhead{\fwhm(\hal)} &\colhead{\fwhm(\hal)} & \colhead{log \lhal} &
\colhead{log(\mbh)} &
\colhead{Obs.} \\
\colhead{} & \colhead{} & \colhead{} & \colhead{(s)} & \colhead{} &
\colhead{(\kms)} & \colhead{(\kms)} & \colhead{(\kms)} &
\colhead{(\kms)} & \colhead{GH07} & \colhead{(\kms)} & \colhead{} &
\colhead{} & \colhead{}}

\startdata
  SDSS J000111.15$-$100155.5 &      & 0.0493 & 5400 & 38 & $~71\pm~3$  & \nodata    & $~82\pm~6$ & $~76\pm~3$ &    1870 & 1660 & 40.11 & 6.00 & m \\
  SDSS J002228.36$-$005830.6 &      & 0.1059 & 5400 & 35 & $~54\pm~5:$ & $~60\pm~5$ & \nodata    & $~57\pm~4$ &    ~807 & ~620 & 41.37 & 5.69 & m \\
  SDSS J004042.10$-$110957.6 &      & 0.0277 & 3600 & 26 & $~57\pm~5$  & $~55\pm~6$ & $~55\pm~5$ & $~55\pm~3$ &    1530 & 2380 & 39.67 & 6.13 & m \\
  SDSS J010712.03+140844.9   & gh01 & 0.0771 & 2982 & 16 & $~38\pm~4$  & $~39\pm~5$ & $~38\pm~9$ & $~38\pm~4$ &    ~914 & ~874 & 41.44 & 6.03 & b \\
  SDSS J011749.81$-$100114.5 &      & 0.1411 & 5400 & 37 & \nodata     & \nodata    & \nodata    & \nodata    &    ~756 & ~640 & 41.53 & 5.79 & m \\
  SDSS J012055.92$-$084945.4 & p    & 0.1246 & 5400 & 14 & $~55\pm~6$  & $~51\pm~6$ & \nodata    & $~53\pm~4$ &    ~700 & 1580 & 40.82 & 6.28 & m \\
  SDSS J014429.16$-$011047.3 &      & 0.0609 & 3600 & 29 & $~68\pm~4$  & \nodata    & $~73\pm12$ & $~70\pm~6$ &    1760 & 2520 & 40.38 & 6.50 & m \\
  SDSS J015804.75$-$005221.9 &      & 0.0807 & 3600 & 34 & $~49\pm~4$  & $~40\pm~3$ & $~47\pm~8$ & $~45\pm~3$ &    1680 & 1180 & 40.56 & 5.90 & m \\
  SDSS J022756.28+005733.0   &      & 0.1280 & 3600 & 16 & \nodata     & \nodata    & \nodata    & \nodata    &    ~929 & 1040 & 41.60 & 6.25 & m \\
  SDSS J022849.51$-$090153.7 &      & 0.0724 & 5400 & 43 & $~69\pm~3$  & \nodata    & $~57\pm10$ & $~63\pm~5$ &    ~697 & ~720 & 40.39 & 5.38 & m \\
  SDSS J023310.79$-$074813.3 & p    & 0.0312 & 1800 & 37 & $101\pm~7$  & $114\pm10$ & $107\pm~4$ & $107\pm~4$ &    1740 & 1800 & 39.92 & 5.99 & m \\
  SDSS J024009.10+010334.5   &      & 0.1956 & 6300 & 16 & \nodata     & \nodata    & \nodata    & \nodata    &    ~737 & ~580 & 41.64 & 5.75 & m \\
  SDSS J024402.24$-$091540.9 &      & 0.1220 & 5400 & 26 & $~73\pm~9:$ & $~79\pm~7$ & \nodata    & $~76\pm~6$ &    ~970 & 1000 & 41.47 & 6.16 & m \\
  SDSS J024912.86$-$081525.6 & gh02 & 0.0297 & 6000 & 22 & $~49\pm~3$  & $~52\pm~2$ & $~60\pm~7$ & $~53\pm~3$ &    ~843 & ~702 & 40.31 & 5.32 & b \\
  SDSS J030417.78+002827.3   &      & 0.0450 & 2400 & 81 & \nodata     & \nodata    & \nodata    & \nodata    &    1000 & ~900 & 41.50 & 6.08 & m \\
  SDSS J032515.59+003408.4   & gh03 & 0.1023 & 3000 & 10 & $~54\pm~6$  & $~47\pm~7$ & \nodata    & $~50\pm~5$ &    ~970 & ~886 & 41.31 & 5.98 & b \\
  SDSS J032707.32$-$075639.3 & p    & 0.1537 & 3600 & 22 & $~75\pm~6$  & \nodata    & \nodata    & $~75\pm~6$ &    ~697 & ~880 & 40.80 & 5.74 & m \\
  SDSS J034745.41+005737.2   &      & 0.1792 & 6600 & 19 & $107\pm11$  & \nodata    & \nodata    & $107\pm11$ &    ~865 & 4717 & 41.55 & 7.58 & m \\
  SDSS J074836.80+182154.2   &      & 0.0715 & 5400 & 28 & $~40\pm~3$  & \nodata    & $~45\pm~8$ & $~42\pm~4$ &    1660 & 1760 & 40.16 & 6.08 & m \\
  SDSS J080629.80+241955.6   &      & 0.0416 & 1200 & 28 & $~73\pm~4$  & $~71\pm~5$ & $~69\pm12$ & $~71\pm~5$ &    ~918 & 1094 & 40.83 & 5.95 & e \\
  SDSS J080907.58+441641.4   &      & 0.0541 & 3600 & 23 & $~66\pm~4$  & $~64\pm~3$ & \nodata    & $~65\pm~3$ &    ~950 & 1104 & 40.78 & 5.94 & b \\
  SDSS J081550.23+250640.9   &      & 0.0726 & 3000 & 17 & $~63\pm~3$  & $~67\pm~4$ & $~65\pm~5$ & $~65\pm~2$ &    ~903 & ~771 & 40.63 & 5.55 & e \\
  SDSS J082325.91+065106.4   & p    & 0.0723 & 3600 & ~9 & $~55\pm~7$  & \nodata    & \nodata    & $~55\pm~7$ &    1320 & 1300 & 39.93 & 5.70 & m \\
  SDSS J082347.95+060636.2   & p    & 0.1037 & 3600 & 13 & $~69\pm~9$  & \nodata    & \nodata    & $~69\pm~9$ &    1480 & 1680 & 40.21 & 6.06 & m \\
  SDSS J082422.21+072550.4   & p    & 0.0815 & 3600 & ~6 & \nodata     & \nodata    & \nodata    & \nodata    &    1220 & ~960 & 41.01 & 5.92 & m \\
  SDSS J082443.28+295923.5   & p    & 0.0254 & 1800 & 47 & $100\pm~7$  & $104\pm~4$ & $118\pm~6$ & $107\pm~3$ &    ~871 & ~691 & 40.35 & 5.33 & e \\
  SDSS J082912.67+500652.3   & gh04 & 0.0436 & 2700 & 35 & $~61\pm~4$  & $~62\pm~3$ & $~58\pm~4$ & $~60\pm~2$ &    ~834 & ~759 & 41.12 & 5.76 & b \\
  SDSS J083346.04+062026.6   &      & 0.1095 & 5400 & 26 & $~39\pm~6$  & $~51\pm~5$ & \nodata    & $~45\pm~4$ &    1070 & ~640 & 40.72 & 5.42 & m \\
  SDSS J083928.45+082102.3   &      & 0.1302 & 3600 & 38 & \nodata     & \nodata    & \nodata    & \nodata    &    ~829 & ~600 & 41.65 & 5.78 & m \\
  SDSS J084011.27+075915.7   & p    & 0.1324 & 3600 & 11 & \nodata     & \nodata    & \nodata    & \nodata    &    1250 & 1060 & 41.09 & 6.04 & m \\
  SDSS J090320.97+045738.0   &      & 0.0567 & 5400 & 126& \nodata     & \nodata    & \nodata    & \nodata    &    ~784 & ~740 & 41.50 & 5.90 & m \\
  SDSS J090431.21+075330.8   &      & 0.0833 & 5400 & 89 & \nodata     & \nodata    & \nodata    & \nodata    &    ~938 & ~860 & 41.29 & 5.94 & m \\
  SDSS J091032.80+040832.4   & p    & 0.0732 & 5400 & 22 & $~72\pm12$  & \nodata    & \nodata    & $~72\pm12$ &    ~864 & ~640 & 40.09 & 5.14 & m \\
  SDSS J091449.05+085321.1   &      & 0.1398 & 5400 & 41 & \nodata     & \nodata    & \nodata    & \nodata    &    ~849 & ~720 & 41.68 & 5.96 & m \\
  SDSS J092547.32+050231.6   &      & 0.1263 & 5400 & 22 & \nodata     & \nodata    & \nodata    & \nodata    &    ~760 & ~580 & 41.55 & 5.71 & m \\
  SDSS J092700.53+084329.4   &      & 0.1124 & 5400 & 26 & $~93\pm12:$ & $107\pm15$ & \nodata    & $100\pm10$ &    1150 & 1220 & 41.29 & 6.26 & m \\
  SDSS J093147.25+063503.2   & p    & 0.0853 & 5400 & 25 & $~52\pm~9$  & \nodata    & \nodata    & $~52\pm~9$ &    ~755 & 1460 & 41.00 & 6.29 & m \\
  SDSS J093147.25+063503.2   & p    & 0.0857 & 3000 & ~8 & $~40\pm~8$  & $~39\pm~9$ & $~28\pm12$ & $~35\pm~6$ &    ~755 & 1360 & 41.00 & 6.22 & e \\
  SDSS J093829.38+034826.6   &      & 0.1193 & 5400 & 28 & $~56\pm~7$  & \nodata    & \nodata    & $~56\pm~7$ &    ~974 & ~800 & 41.21 & 5.84 & m \\
  SDSS J094057.19+032401.2   &      & 0.0606 & 5400 & 210& $~82\pm~3$  & \nodata    & \nodata    & $~82\pm~3$ &    ~908 & ~800 & 41.45 & 5.95 & m \\
  SDSS J094310.12+604559.1   & gh05 & 0.0743 & 3600 & 16 & \nodata     & \nodata    & \nodata    & \nodata    &    ~807 & ~679 & 41.34 & 5.76 & b \\
  SDSS J094529.36+093610.4   &      & 0.0131 & 3600 & 81 & $~75\pm~2$  & $~77\pm~3$ & \nodata    & $~76\pm~2$ &    1930 & 1720 & 40.53 & 6.22 & m \\
  SDSS J095151.82+060143.7   &      & 0.0932 & 3600 & 35 & $~70\pm~6$  & $~83\pm10$ & \nodata    & $~76\pm~6$ &    1260 & ~660 & 41.01 & 5.58 & m \\
  SDSS J100035.47+052428.5   &      & 0.0785 & 2100 & 29 & \nodata     & \nodata    & \nodata    & \nodata    &    1000 & ~940 & 41.62 & 6.17 & m \\
  SDSS J101108.40+002908.7   & gh06 & 0.1002 & 1800 & 10 & $~50\pm~7$  & $~61\pm~7$ & \nodata    & $~55\pm~5$ &    1010 & 1083 & 41.36 & 6.18 & b \\
  SDSS J101627.32$-$000714.5 & gh07 & 0.0950 & 3600 & 15 & \nodata     & $~55\pm~7$ & \nodata    & $~55\pm~7$ & \nodata & ~633 & 41.06 & 5.57 & b \\
  SDSS J102124.87+012720.3   &      & 0.0668 & 3600 & 36 & $~78\pm~3$  & \nodata    & \nodata    & $~78\pm~3$ &    1690 & 1600 & 40.39 & 6.09 & m \\
  SDSS J102348.44+040553.7   & p    & 0.0988 & 5400 & 27 & $~91\pm13$  & \nodata    & \nodata    & $~91\pm13$ &    ~696 & ~520 & 40.95 & 5.34 & m \\
  SDSS J103518.74+073406.2   &      & 0.0674 & 5400 & 55 & $109\pm~4$  & \nodata    & \nodata    & $109\pm~4$ &    ~867 & ~800 & 41.27 & 5.87 & m \\
  SDSS J104210.03$-$001814.7 &      & 0.1144 & 5400 & 32 & \nodata     & \nodata    & \nodata    & \nodata    &    ~816 & ~800 & 41.64 & 6.04 & m \\
  SDSS J105755.66+482502.0   &      & 0.0732 & 1200 & 22 & $~50\pm~3$  & $~49\pm~4$ & $~38\pm~3$ & $~45\pm~2$ &    ~957 & ~898 & 40.62 & 5.68 & e \\
  SDSS J110501.97+594103.6   &      & 0.0337 & ~900 & 50 & $124\pm~5$  & $125\pm~8$ & $120\pm19$ & $123\pm~7$ &    ~702 & 3672 & 40.51 & 6.89 & e \\
  SDSS J110540.46+035309.0   &      & 0.0993 & 3600 & 47 & \nodata     & \nodata    & \nodata    & \nodata    &    ~820 & ~800 & 41.62 & 6.03 & m \\
  SDSS J111031.61+022043.2   &      & 0.0799 & 5400 & 92 & $~78\pm~3$  & $~76\pm~4$ & \nodata    & $~77\pm~3$ &    1000 & ~920 & 41.39 & 6.05 & m \\
  SDSS J111749.17+044315.5   &      & 0.1082 & 5400 & 36 & $~62\pm~9:$ & $~76\pm~6$ & \nodata    & $~69\pm~5$ &    ~826 & ~680 & 41.38 & 5.77 & m \\
  SDSS J112526.51+022039.0   &      & 0.0490 & 5400 & 136& $~91\pm~2$  & \nodata    & $~83\pm10$ & $~87\pm~5$ &    1090 & ~940 & 41.14 & 5.96 & m \\
  SDSS J114339.49$-$024316.3 &      & 0.0937 & 5400 & 61 & $~97\pm~8$  & $~98\pm~5$ & \nodata    & $~97\pm~5$ &    ~919 & 1460 & 41.18 & 6.37 & m \\
  SDSS J114343.76+550019.3   & p    & 0.0272 & 1800 & 22 & $~31\pm~3$  & $~31\pm~3$ & $~32\pm~3$ & $~31\pm~2$ &    1070 & 1393 & 40.39 & 5.97 & e \\
  SDSS J114439.34+025506.5   & p    & 0.1018 & 5400 & 21 & $~48\pm~4$  & $~47\pm~6$ & \nodata    & $~47\pm~4$ &    ~942 & ~720 & 40.51 & 5.43 & m \\
  SDSS J114633.98+100244.9   &      & 0.1245 & 5400 & 31 & \nodata     & $~62\pm~8$ & \nodata    & $~62\pm~8$ &    ~790 & ~780 & 41.54 & 5.97 & m \\
  SDSS J115138.24+004946.4   & gh09 & 0.1950 & 1200 & 13 & \nodata     & \nodata    & \nodata    & \nodata    &    ~810 & 1304 & 41.74 & 6.52 & e \\
  SDSS J121518.23+014751.1   &      & 0.0713 & 5400 & 95 & $~74\pm~4$  & \nodata    & $~88\pm~4$ & $~81\pm~3$ &    ~910 & 1000 & 41.08 & 5.99 & m \\
  SDSS J122342.81+581446.1   &      & 0.0146 & 1800 & 29 & $~44\pm~3$  & $~47\pm~3$ & $~46\pm~3$ & $~45\pm~2$ &    ~979 & 1577 & 40.30 & 6.04 & e \\
  SDSS J124035.81$-$002919.4 & gh10 & 0.0812 & 1800 & 26 & $~49\pm~5$  & $~63\pm~6$ & $~58\pm~7$ & $~56\pm~3$ &    ~915 & ~713 & 41.64 & 5.93 & b \\
  SDSS J125055.28$-$015556.6 & gh11 & 0.0815 & 1800 & 18 & $~68\pm~5$  & $~68\pm~8$ & $~64\pm~8$ & $~66\pm~4$ & \nodata & 2266 & 41.27 & 6.80 & b \\
  SDSS J131310.12+051942.1   &      & 0.0492 & 5400 & 76 & $~68\pm~3$  & \nodata    & $~80\pm~4$ & $~74\pm~3$ &    ~888 & ~580 & 40.68 & 5.32 & m \\
  SDSS J131310.12+051942.1   &      & 0.0489 & 1200 & 16 & $~64\pm~5$  & $~65\pm~5$ & $~68\pm~6$ & $~65\pm~3$ &    ~888 & ~645 & 40.68 & 5.41 & e \\
  SDSS J131651.29+055646.9   &      & 0.0554 & 5400 & 101& $~82\pm~3$  & \nodata    & \nodata    & $~82\pm~3$ &    1260 & ~980 & 41.04 & 5.95 & m \\
  SDSS J131659.37+035319.8   & p    & 0.0459 & 5400 & 36 & $~82\pm~6$  & \nodata    & $~80\pm12$ & $~81\pm~7$ &    ~887 & ~880 & 40.61 & 5.66 & m \\
  SDSS J131926.52+105610.9   &      & 0.0647 & 5400 & 174& $~47\pm~3$  & \nodata    & \nodata    & $~47\pm~3$ &    ~840 & ~860 & 41.10 & 5.86 & m \\
  SDSS J134144.51$-$005832.9 &      & 0.1476 & 5400 & 31 & \nodata     & \nodata    & \nodata    & \nodata    &    ~835 & ~660 & 41.39 & 5.75 & m \\
  SDSS J141234.67$-$003500.0 & gh13 & 0.1269 & 1800 & 16 & \nodata     & \nodata    & \nodata    & \nodata    &    ~884 & ~945 & 41.56 & 6.15 & b \\
  SDSS J143450.62+033842.5   & gh14 & 0.0284 & 1800 & 17 & $~46\pm~6$  & $~63\pm~3$ & $~63\pm~6$ & $~57\pm~3$ &    1050 & 1001 & 40.34 & 5.65 & b \\
  SDSS J144052.60$-$023506.2 &      & 0.0448 & 5400 & 143& \nodata     & \nodata    & $~73\pm~8$ & $~73\pm~8$ &    ~950 & ~840 & 41.21 & 5.89 & m \\
  SDSS J144705.46+003653.2   &      & 0.0953 & 5400 & 44 & $~63\pm~4$  & $~65\pm~7$ & \nodata    & $~64\pm~4$ &    1160 & 1500 & 40.96 & 6.29 & m \\
  SDSS J145045.54$-$014752.8 &      & 0.0996 & 5400 & 61 & $131\pm~6$  & $145\pm10$ & \nodata    & $138\pm~6$ &    ~955 & 2420 & 41.49 & 6.96 & m \\
  SDSS J150754.38+010816.7   &      & 0.0613 & 5400 & 114& $131\pm~5$  & \nodata    & $133\pm~3$ & $132\pm~3$ &    ~699 & 1680 & 40.40 & 6.14 & m \\
  SDSS J153425.59+040806.7   &      & 0.0395 & 2200 & ~7 & \nodata     & \nodata    & \nodata    & \nodata    &    ~927 & ~449 & 40.02 & 4.79 & e \\
  SDSS J154257.49+030653.2   &      & 0.0655 & 5400 & 117& \nodata     & \nodata    & \nodata    & \nodata    &    1070 & ~980 & 41.41 & 6.12 & m \\
  SDSS J155005.95+091035.7   &      & 0.0916 & 5400 & 51 & $~80\pm10$  & $~77\pm~8$ & \nodata    & $~78\pm~6$ &    ~835 & 1100 & 41.10 & 6.08 & m \\
  SDSS J161227.84+010159.7   &      & 0.0973 & 7200 & 70 & \nodata     & \nodata    & \nodata    & \nodata    &    ~944 & ~920 & 41.40 & 6.05 & m \\
  SDSS J161751.98$-$001957.4 &      & 0.0573 & 5100 & 14 & $~65\pm~6$  & \nodata    & \nodata    & $~65\pm~6$ &    1120 & 1020 & 40.37 & 5.68 & m \\
  SDSS J162403.63$-$005410.3 &      & 0.0468 & 5400 & 35 & $~94\pm~4$  & \nodata    & \nodata    & $~94\pm~4$ &    1150 & 1340 & 40.64 & 6.05 & m \\
  SDSS J162636.40+350242.0   &      & 0.0342 & 1800 & 34 & $~48\pm~2$  & $~53\pm~2$ & $~56\pm~2$ & $~52\pm~1$ &    ~802 & ~714 & 40.76 & 5.54 & e \\
  SDSS J163159.59+243740.2   &      & 0.0436 & 2900 & 38 & $~63\pm~3$  & $~61\pm~3$ & $~74\pm~4$ & $~66\pm~2$ &    ~839 & ~541 & 40.84 & 5.33 & e \\
  SDSS J170246.09+602818.9   & gh16 & 0.0692 & 1100 & 18 & \nodata     & $~81\pm11$ & $~82\pm~9$ & $~81\pm~7$ & \nodata & 1116 & 41.25 & 6.16 & b \\
  SDSS J172759.15+542147.0   & gh17 & 0.0997 & 3500 & 17 & \nodata     & $~67\pm~8$ & \nodata    & $~67\pm~8$ &    ~806 & ~656 & 41.27 & 5.68 & b \\
  SDSS J205822.14$-$065004.3 &      & 0.0742 & 5400 & 62 & $~58\pm~3:$ & \nodata    & \nodata    & $~58\pm~3$ &    ~917 & ~860 & 41.54 & 6.06 & m \\
  SDSS J213728.62$-$083823.3 &      & 0.1609 & 7200 & 36 & \nodata     & \nodata    & \nodata    & \nodata    &    ~904 & 1000 & 41.76 & 6.29 & m \\
  SDSS J215658.30+110343.1   &      & 0.1081 & 3600 & 36 & $~88\pm~8$  & \nodata    & \nodata    & $~88\pm~8$ & \nodata & ~990 & 42.14 & 6.45 & b \\
  SDSS J221139.16$-$010535.0 &      & 0.0925 & 7200 & 20 & $~69\pm10$  & $~68\pm11$ & \nodata    & $~68\pm~7$ &    1070 & 1540 & 40.95 & 6.31 & m \\
  SDSS J230649.77+005023.4   &      & 0.0610 & 3600 & 39 & $~58\pm~3$  & \nodata    & $~73\pm~6$ & $~65\pm~3$ &    1800 & 1479 & 40.55 & 6.10 & m \\
  SDSS J232159.06+000738.8   & gh18 & 0.1840 & 3600 & 12 & $~76\pm~9$  & \nodata    & \nodata    & $~76\pm~9$ & \nodata & 1531 & 41.52 & 6.56 & b \\
  SDSS J233837.10$-$002810.3 & gh19 & 0.0357 & 3600 & 26 & $~58\pm~3$  & $~56\pm~2$ & $~55\pm~3$ & $~56\pm~2$ & \nodata & 1553 & 40.06 & 5.92 & b \\
  SDSS J234807.14$-$091202.6 &      & 0.0779 & 5400 & 34 & $~80\pm~7$  & \nodata    & \nodata    & $~80\pm~7$ &    1490 & 1480 & 40.76 & 6.19 & m \\

\enddata

\tablecomments{Object names are given by their SDSS coordinate
designations.  Those names in \citet{GH04} are also given.  Objects
with ``possible'' broad \hal\ are indicated as ``p''.  $S/N$ is the
mean signal-to-noise ratio per pixel in the spectrum in the spectral
region around or redward of \mgb\ used to measure \sigmastar.  For
\mage\ data, if the \mgb\ fitting region and Fe fitting region are
on the same echelle order, we measure \sigmastar\ from one region
extending across these two, and list the result as $\sigma$(\mgb).
The \sigmastar\ with ``:'' is dubious due to limited numbers of
fitting templates.  The \fwhm(\hal)$_{GH07}$ and \lhal\ values are
from GH07 if available and otherwise from \citet{GH04}. For SDSS
J215658.30+110343.1, we use \lctm\ from \citet{barth05} and the
\lctm-\lhal\ relation in \citet{GH05b} to estimate \lhal. Virial
mass estimates of \mbh\ are in units of \msun.  The observational
sub-samples are indicated as \textit{b}, \textit{e} and \textit{m}
for \esi\ sample in \citet{barth05}, new \esi\ observation and new
\mage\ observation, respectively. } \label{table1}
\end{deluxetable}

\begin{deluxetable}{lrcccc}
\tablewidth{0pt} \tabletypesize{\scriptsize}
\tablecaption{Regression
parameters} \tablehead{ \colhead{Sample} &  \colhead{N}   &
\colhead{Uncert.} & \colhead{$\alpha$} & \colhead{$\beta$} &
\colhead{$\epsilon_0$}
\\
\colhead{(1)} & \colhead{(2)} & \colhead{(3)} & \colhead{(4)} &
\colhead{(5)} & \colhead{(6)} } \tablecolumns{6}
\startdata

Full             & 155  & 1$\sigma$ & $7.68\pm0.08$ &  $3.32\pm0.22$ & $0.46\pm0.03$ \\
Full             & 155  & 3$\sigma$ & $7.69\pm0.08$ &  $3.28\pm0.22$ & $0.28\pm0.05$ \\
Full w/o ``p''   & 142  & 1$\sigma$ & $7.75\pm0.08$ &  $3.48\pm0.21$ & $0.41\pm0.03$ \\

\enddata
\label{tab:reg} \tablecomments{ Col. (1): Data set considered; the
``Full'' sample comprises objects with both \mbh\ and \sigmastar\
available, including the SDSS sample and 24 RM AGNs with stellar
velocity dispersion measurements presented by \citet{woo2010}, NGC
4395 and POX 52. The ``Full w/o p'' subsample denotes the full
sample minus the objects classified as having possible broad \hal.
Col. (2): Number of objects in each sample. Col. (3): Adopted
uncertainties for \mbh. Col. (4): Zero point assuming $\log \mbh\ =
\alpha\ + \beta\ \log (\sigmastar/200~\mathrm{km~s^{-1}})$, for
$f=0.75$.  Col. (5): \msigma\ slope.  Col. (6): intrinsic scatter.}
\end{deluxetable}

\begin{deluxetable}{lrccccccc}
\tablewidth{0pt} \tabletypesize{\scriptsize}
\tablecaption{Regression parameters for barred / unbarred disk
galaxies} \tablehead{ \colhead{} &  \colhead{} &
\multicolumn{3}{c}{Free fit} & \colhead{} & \multicolumn{3}{c}{Fit
with fixed slope}
\\
\cline{3-5} \cline{7-9} \\
\colhead{Sample} &  \colhead{N} & \colhead{$\alpha$} &
\colhead{$\beta$} & \colhead{$\epsilon_0$} & \colhead{} &
\colhead{$\alpha$} & \colhead{$\beta$} & \colhead{$\epsilon_0$}
\\
\colhead{(1)} & \colhead{(2)} & \colhead{(3)} & \colhead{(4)} &
\colhead{(5)} & \colhead{} & \colhead{(6)} & \colhead{(7)} &
\colhead{(8)} } \tablecolumns{8}
\startdata

Disk               & 63 & $7.50\pm0.13$ &  $3.04\pm0.30$ & $0.43\pm0.05$ & & $7.96\pm0.07$ &  $4.24$ & $0.50$ \\
Barred             & 25 & $7.81\pm0.27$ &  $4.13\pm0.72$ & $0.46\pm0.09$ & & $7.86\pm0.09$ &  $4.24$ & $0.40$ \\
Unbarred           & 38 & $7.44\pm0.15$ &  $2.79\pm0.34$ & $0.42\pm0.06$ & & $8.04\pm0.09$ &  $4.24$ & $0.54$ \\
Disk w/o ``p''     & 56 & $7.56\pm0.13$ &  $3.19\pm0.31$ & $0.42\pm0.05$ & & $7.95\pm0.07$ &  $4.24$ & $0.46$ \\
Barred w/o ``p''   & 22 & $7.80\pm0.29$ &  $4.03\pm0.80$ & $0.46\pm0.09$ & & $7.87\pm0.09$ &  $4.24$ & $0.40$ \\
Unbarred w/o ``p'' & 34 & $7.52\pm0.15$ &  $3.01\pm0.33$ & $0.40\pm0.06$ & & $8.01\pm0.09$ &  $4.24$ & $0.48$ \\

\enddata
\label{tab:bar} \tablecomments{ Col. (1): Data set considered; the
``Disk'' (``Barred'' / ``Unbarred'')  sample comprises the ``Full''
sample with morphological type available and classified as (barred /
unbarred) disk galaxies. The ``Disk w/o p'' subsample denotes the
``Disk'' sample minus the objects classified as having ``possible''
broad \hal. Col. (2): Number of objects in each subsample. Col. (3):
Zeropoint assuming $\log \mbh\ = \alpha\ + \beta\ \log
(\sigmastar/200~\mathrm{km~s^{-1}})$, for $f=0.75$. Col. (4):
\msigma\ slope.  Col. (5): Intrinsic scatter. Col. (6): Zeropoint
of \msigma\ relation with the slope fixed to that of inactive
galaxies listed in Col.(7), and the upper limits on intrinsic
scatter listed in Col. (8). }
\end{deluxetable}

\begin{deluxetable}{lrcccccccc}
\tablewidth{0pt} \tabletypesize{\scriptsize}
\tablecaption{Regression
parameters for different disk inclination} \tablehead{ \colhead{} &
\colhead{} & \colhead{} & \multicolumn{3}{c}{Free fit} & \colhead{}
& \multicolumn{3}{c}{Fit with fixed slope}
\\
\cline{4-6} \cline{8-10} \\
\colhead{Sample} & \colhead{axis ratio} & \colhead{N} &
\colhead{$\alpha$} & \colhead{$\beta$} & \colhead{$\epsilon_0$} &
\colhead{} & \colhead{$\alpha$} & \colhead{$\beta$} &
\colhead{$\epsilon_0$}
\\
\colhead{(1)} & \colhead{(2)} & \colhead{(3)} & \colhead{(4)} &
\colhead{(5)} & \colhead{(6)} & \colhead{} & \colhead{(7)} &
\colhead{(8)} & \colhead{(9)} } \tablecolumns{9} \startdata

High   & $     b/a<0.72$ & 20 & $7.42\pm0.25$ &  $3.24\pm0.70$ & $0.50\pm0.10$ & & $7.74\pm0.11$ &  $4.24$ & $0.48$ \\
Medium & $0.72<b/a<0.88$ & 20 & $7.70\pm0.20$ &  $3.86\pm0.49$ & $0.38\pm0.10$ & & $7.85\pm0.09$ &  $4.24$ & $0.36$ \\
Low    & $     b/a>0.88$ & 15 & $7.59\pm0.32$ &  $2.76\pm0.67$ & $0.44\pm0.12$ & & $8.25\pm0.13$ &  $4.24$ & $0.48$ \\

\enddata
\label{tab:inc} \tablecomments{ Col. (1): Data set considered; the
``High'' (``Medium'' / ``Low'')  sample comprises the ``Disk''
sample with the axis ratio of the disk component available, divided
into three bins of inclination angle.  Col. (2): axis ratio range of
each subsample.  Col. (3): Number of objects in each subsample. Col.
(4): Zeropoint assuming $\log \mbh\ = \alpha\ + \beta\ \log
(\sigmastar/200~\mathrm{km~s^{-1}})$, for $f=0.75$. Col. (5):
\msigma\ slope.  Col. (6): Intrinsic scatter. Col. (7): Zeropoint
of \msigma\ relation with the slope fixed to that of inactive
galaxies listed in Col.(8), and the upper limits on intrinsic
scatter listed in Col. (9). }
\end{deluxetable}

\end{document}